\documentclass[%
 superscriptaddress,
 reprint,
 showpacs,preprintnumbers,
 nofootinbib,
 amsmath,amssymb,
 aps,
 prd,
 floatfix,
]{revtex4-1}

\usepackage{graphicx}

\usepackage{xcolor}
\usepackage{lineno}
\usepackage{float}
\usepackage{afterpage}

\usepackage{parskip}
\usepackage{lipsum}

\usepackage{amssymb,amsmath,amstext,amsthm,amsfonts,amsfonts}
\usepackage[utf8]{inputenc}
\usepackage[linesnumbered,ruled]{algorithm2e}
\usepackage{url}
\usepackage{soul}
\usepackage{bm}
\usepackage[normalem]{ulem}
\usepackage{multirow, makecell}
\usepackage{booktabs}

\makeatletter
\def\@fnsymbol#1{\ensuremath{\ifcase#1\or
   *\or \dagger\or \ddagger\or \mathsection\or \mathparagraph\or \|\or
   **\or \dagger\dagger\or \ddagger\ddagger\or
   \mathsection\mathsection\or \mathparagraph\mathparagraph\or \|\|\or
   ***\or \dagger\dagger\dagger\or \ddagger\ddagger\ddagger\or
   \mathsection\mathsection\mathsection\or \mathparagraph\mathparagraph\mathparagraph\or \|\|\|\or
   ****\or \dagger\dagger\dagger\dagger\or \ddagger\ddagger\ddagger\ddagger
   \else\@ctrerr\fi}}
\makeatother

\usepackage{hyperref}
\hypersetup{
    hidelinks,               
    colorlinks=true,         
    citecolor=teal,          
    linkcolor=teal,           
    urlcolor=teal           
}

\begin{document}

\title{Signal selection and model-independent extraction of the neutrino neutral-current single $\mathbf{\pi}^+$ cross section with the T2K experiment}


\newcommand{\INSTHD}{\affiliation{University Autonoma Madrid, Department of Theoretical Physics, 28049 Madrid, Spain}}
\newcommand{\INSTFE}{\affiliation{Boston University, Department of Physics, Boston, Massachusetts, U.S.A.}}
\newcommand{\INSTD}{\affiliation{University of British Columbia, Department of Physics and Astronomy, Vancouver, British Columbia, Canada}}
\newcommand{\INSTGA}{\affiliation{University of California, Irvine, Department of Physics and Astronomy, Irvine, California, U.S.A.}}
\newcommand{\INSTI}{\affiliation{IRFU, CEA, Universit\'e Paris-Saclay, F-91191 Gif-sur-Yvette, France}}
\newcommand{\INSTGB}{\affiliation{University of Colorado at Boulder, Department of Physics, Boulder, Colorado, U.S.A.}}
\newcommand{\INSTFH}{\affiliation{Duke University, Department of Physics, Durham, North Carolina, U.S.A.}}
\newcommand{\INSTJA}{\affiliation{E\"{o}tv\"{o}s Lor\'{a}nd University, Department of Atomic Physics, Budapest, Hungary}}
\newcommand{\INSTEF}{\affiliation{ETH Zurich, Institute for Particle Physics and Astrophysics, Zurich, Switzerland}}
\newcommand{\INSTIG}{\affiliation{VNU University of Science, Vietnam National University, Hanoi, Vietnam}}
\newcommand{\INSTIE}{\affiliation{CERN European Organization for Nuclear Research, CH-1211 Gen\'eve 23, Switzerland}}
\newcommand{\INSTEG}{\affiliation{University of Geneva, Section de Physique, DPNC, Geneva, Switzerland}}
\newcommand{\INSTHJ}{\affiliation{University of Glasgow, School of Physics and Astronomy, Glasgow, United Kingdom}}
\newcommand{\INSTJG}{\affiliation{Ghent University, Department of Physics and Astronomy, Proeftuinstraat 86, B-9000 Gent, Belgium}}
\newcommand{\INSTDG}{\affiliation{H. Niewodniczanski Institute of Nuclear Physics PAN, Cracow, Poland}}
\newcommand{\INSTCB}{\affiliation{High Energy Accelerator Research Organization (KEK), Tsukuba, Ibaraki, Japan}}
\newcommand{\INSTIB}{\affiliation{University of Houston, Department of Physics, Houston, Texas, U.S.A.}}
\newcommand{\INSTED}{\affiliation{Institut de Fisica d'Altes Energies (IFAE) - The Barcelona Institute of Science and Technology, Campus UAB, Bellaterra (Barcelona) Spain}}
\newcommand{\INSTJC}{\affiliation{Institut f\"ur Physik, Johannes Gutenberg-Universit\"at Mainz, Staudingerweg 7, 55128 Mainz, Germany}}
\newcommand{\INSTHH}{\affiliation{Institute For Interdisciplinary Research in Science and Education (IFIRSE), ICISE, Quy Nhon, Vietnam}}
\newcommand{\INSTEI}{\affiliation{Imperial College London, Department of Physics, London, United Kingdom}}
\newcommand{\INSTGF}{\affiliation{INFN Sezione di Bari and Universit\`a e Politecnico di Bari, Dipartimento Interuniversitario di Fisica, Bari, Italy}}
\newcommand{\INSTBE}{\affiliation{INFN Sezione di Napoli and Universit\`a di Napoli, Dipartimento di Fisica, Napoli, Italy}}
\newcommand{\INSTBF}{\affiliation{INFN Sezione di Padova and Universit\`a di Padova, Dipartimento di Fisica, Padova, Italy}}
\newcommand{\INSTBD}{\affiliation{INFN Sezione di Roma and Universit\`a di Roma ``La Sapienza'', Roma, Italy}}
\newcommand{\INSTEB}{\affiliation{Institute for Nuclear Research of the Russian Academy of Sciences, Moscow, Russia}}
\newcommand{\INSTHI}{\affiliation{International Centre of Physics, Institute of Physics (IOP), Vietnam Academy of Science and Technology (VAST), 10 Dao Tan, Ba Dinh, Hanoi, Vietnam}}
\newcommand{\INSTJD}{\affiliation{ILANCE, CNRS – University of Tokyo International Research Laboratory, Kashiwa, Chiba 277-8582, Japan}}
\newcommand{\INSTHA}{\affiliation{Kavli Institute for the Physics and Mathematics of the Universe (WPI), The University of Tokyo Institutes for Advanced Study, University of Tokyo, Kashiwa, Chiba, Japan}}
\newcommand{\INSTID}{\affiliation{Keio University, Department of Physics, Kanagawa, Japan}}
\newcommand{\INSTIF}{\affiliation{King's College London, Department of Physics, Strand, London WC2R 2LS, United Kingdom}}
\newcommand{\INSTCC}{\affiliation{Kobe University, Kobe, Japan}}
\newcommand{\INSTCD}{\affiliation{Kyoto University, Department of Physics, Kyoto, Japan}}
\newcommand{\INSTEJ}{\affiliation{Lancaster University, Physics Department, Lancaster, United Kingdom}}
\newcommand{\INSTII}{\affiliation{Lawrence Berkeley National Laboratory, Berkeley, California, U.S.A.}}
\newcommand{\INSTBA}{\affiliation{Ecole Polytechnique, IN2P3-CNRS, Laboratoire Leprince-Ringuet, Palaiseau, France}}
\newcommand{\INSTFC}{\affiliation{University of Liverpool, Department of Physics, Liverpool, United Kingdom}}
\newcommand{\INSTFI}{\affiliation{Louisiana State University, Department of Physics and Astronomy, Baton Rouge, Louisiana, U.S.A.}}
\newcommand{\INSTIH}{\affiliation{Joint Institute for Nuclear Research, Dubna, Moscow Region, Russia}}
\newcommand{\INSTHB}{\affiliation{Michigan State University, Department of Physics and Astronomy,  East Lansing, Michigan, U.S.A.}}
\newcommand{\INSTCE}{\affiliation{Miyagi University of Education, Department of Physics, Sendai, Japan}}
\newcommand{\INSTDF}{\affiliation{National Centre for Nuclear Research, Warsaw, Poland}}
\newcommand{\INSTFJ}{\affiliation{State University of New York at Stony Brook, Department of Physics and Astronomy, Stony Brook, New York, U.S.A.}}
\newcommand{\INSTEH}{\affiliation{STFC, Rutherford Appleton Laboratory, Harwell Oxford,  and  Daresbury Laboratory, Warrington, United Kingdom}}
\newcommand{\INSTGJ}{\affiliation{Okayama University, Department of Physics, Okayama, Japan}}
\newcommand{\INSTCF}{\affiliation{Osaka Metropolitan University, Department of Physics, Osaka, Japan}}
\newcommand{\INSTGG}{\affiliation{Oxford University, Department of Physics, Oxford, United Kingdom}}
\newcommand{\INSTIC}{\affiliation{University of Pennsylvania, Department of Physics and Astronomy,  Philadelphia, Pennsylvania, U.S.A.}}
\newcommand{\INSTGC}{\affiliation{University of Pittsburgh, Department of Physics and Astronomy, Pittsburgh, Pennsylvania, U.S.A.}}
\newcommand{\INSTGD}{\affiliation{University of Rochester, Department of Physics and Astronomy, Rochester, New York, U.S.A.}}
\newcommand{\INSTHC}{\affiliation{Royal Holloway University of London, Department of Physics, Egham, Surrey, United Kingdom}}
\newcommand{\INSTBC}{\affiliation{RWTH Aachen University, III. Physikalisches Institut, Aachen, Germany}}
\newcommand{\INSTJF}{\affiliation{School of Physics and Astronomy, University of Minnesota, Minneapolis, Minnesota, U.S.A.}}
\newcommand{\INSTJB}{\affiliation{Departamento de F\'isica At\'omica, Molecular y Nuclear, Universidad de Sevilla, 41080 Sevilla, Spain}}
\newcommand{\INSTFB}{\affiliation{University of Sheffield, School of Mathematical and Physical Sciences, Sheffield, United Kingdom}}
\newcommand{\INSTDI}{\affiliation{University of Silesia, Institute of Physics, Katowice, Poland}}
\newcommand{\INSTIA}{\affiliation{SLAC National Accelerator Laboratory, Stanford University, Menlo Park, California, U.S.A.}}
\newcommand{\INSTBB}{\affiliation{Sorbonne Universit\'e, CNRS/IN2P3, Laboratoire de Physique Nucl\'eaire et de Hautes Energies (LPNHE), Paris, France}}
\newcommand{\INSTJE}{\affiliation{South Dakota School of Mines and Technology, 501 East Saint Joseph Street, Rapid City, SD 57701, United States}}
\newcommand{\INSTCH}{\affiliation{University of Tokyo, Department of Physics, Tokyo, Japan}}
\newcommand{\INSTBJ}{\affiliation{University of Tokyo, Institute for Cosmic Ray Research, Kamioka Observatory, Kamioka, Japan}}
\newcommand{\INSTCG}{\affiliation{University of Tokyo, Institute for Cosmic Ray Research, Research Center for Cosmic Neutrinos, Kashiwa, Japan}}
\newcommand{\INSTHF}{\affiliation{Institute of Science Tokyo, Department of Physics, Tokyo}}
\newcommand{\INSTGI}{\affiliation{Tokyo Metropolitan University, Department of Physics, Tokyo, Japan}}
\newcommand{\INSTHG}{\affiliation{Tokyo University of Science, Faculty of Science and Technology, Department of Physics, Noda, Chiba, Japan}}
\newcommand{\INSTB}{\affiliation{TRIUMF, Vancouver, British Columbia, Canada}}
\newcommand{\INSTJH}{\affiliation{University of Toyama, Department of Physics, Toyama, Japan}}
\newcommand{\INSTDJ}{\affiliation{University of Warsaw, Faculty of Physics, Warsaw, Poland}}
\newcommand{\INSTDH}{\affiliation{Warsaw University of Technology, Institute of Radioelectronics and Multimedia Technology, Warsaw, Poland}}
\newcommand{\INSTIJ}{\affiliation{Tohoku University, Faculty of Science, Department of Physics, Miyagi, Japan}}
\newcommand{\INSTFD}{\affiliation{University of Warwick, Department of Physics, Coventry, United Kingdom}}
\newcommand{\INSTEA}{\affiliation{Wroclaw University, Faculty of Physics and Astronomy, Wroclaw, Poland}}
\newcommand{\INSTHE}{\affiliation{Yokohama National University, Department of Physics, Yokohama, Japan}}
\newcommand{\INSTH}{\affiliation{York University, Department of Physics and Astronomy, Toronto, Ontario, Canada}}

\INSTHD
\INSTFE
\INSTD
\INSTGA
\INSTI
\INSTGB
\INSTFH
\INSTJA
\INSTEF
\INSTIG
\INSTIE
\INSTEG
\INSTHJ
\INSTJG
\INSTDG
\INSTCB
\INSTIB
\INSTED
\INSTJC
\INSTHH
\INSTEI
\INSTGF
\INSTBE
\INSTBF
\INSTBD
\INSTEB
\INSTHI
\INSTJD
\INSTHA
\INSTID
\INSTIF
\INSTCC
\INSTCD
\INSTEJ
\INSTII
\INSTBA
\INSTFC
\INSTFI
\INSTIH
\INSTHB
\INSTCE
\INSTDF
\INSTFJ
\INSTEH
\INSTGJ
\INSTCF
\INSTGG
\INSTIC
\INSTGC
\INSTGD
\INSTHC
\INSTBC
\INSTJF
\INSTJB
\INSTFB
\INSTDI
\INSTIA
\INSTBB
\INSTJE
\INSTCH
\INSTBJ
\INSTCG
\INSTHF
\INSTGI
\INSTHG
\INSTB
\INSTJH
\INSTDJ
\INSTDH
\INSTIJ
\INSTFD
\INSTEA
\INSTHE
\INSTH

\author{K.\,Abe}\INSTBJ
\author{S.\,Abe}\INSTBJ
\author{R.\,Akutsu}\INSTCB
\author{H.\,Alarakia-Charles}\INSTEJ
\author{Y.I.\,Alj Hakim}\INSTFB
\author{S.\,Alonso Monsalve}\INSTEF
\author{L.\,Anthony}\INSTEI
\author{S.\,Aoki}\INSTCC
\author{K.A.\,Apte}\INSTEI
\author{T.\,Arai}\INSTCH
% ...

\collaboration{The T2K Collaboration}\noaffiliation
    
\begin{abstract}
\noindent This article presents a study of single $\pi^+$ production in neutrino neutral-current interactions (NC1$\pi^+$) using the FGD1 hydrocarbon target of the ND280 detector of the T2K experiment. We report the largest sample of such events selected by any experiment, providing the first new data for this channel in over four decades and the first using a sub-GeV neutrino flux. The signal selection strategy and its performance are detailed together with validations of a robust cross section extraction methodology. The measured flux-averaged integrated cross-section is $ \sigma = (6.07 \pm 1.22 )\times 10^{-41} \,\, \text{cm}^2/\text{nucleon}$, 1.3~$\sigma~$ above the NEUT v5.4.0 expectation.

\end{abstract}

\maketitle

\section{Introduction}

The observation of neutrino oscillations~\cite{Super-Kamiokande:1998kpq,SNO:2002tuh} has boosted neutrino research in the last two decades, focusing on measurements of three-flavor mixing parameters~\cite{Bilenky:1998dt, Gonzalez-Garcia:2007dlo, Esteban:2020cvm, ParticleDataGroup:2022pth} and searches for exotic physics. Better understanding neutrino interactions has become a priority, particularly at energies of $E_\nu\sim1$~GeV, where nuclear effects involved in neutrino-nucleus interactions pose a significant experimental challenge~\cite{NuSTEC:2017hzk}. Despite an increasing abundance of inclusive and exclusive cross section results, some channels still remain largely unexplored experimentally. A prime example of this is single $\pi^+$ production in neutrino-nucleus neutral-current interactions, 
\begin{equation}
\nu+p \rightarrow \nu + n + \pi^+,
\end{equation}
in which a neutrino scattering off a proton is able to transform it into a neutron while additionally producing a positive pion. Existing data for this channel consists of observations in bubble chamber experiments built in the 1970s to test the validity of the Standard Model, which predicted the existence of this process. In particular, the Gargamelle propane freon bubble chamber~\cite{GargamelleNeutrinoPropane:1977hya} and the ANL deuteron bubble chamber~\cite{Derrick:1980nr} reported the observation of tens of interactions associated to this channel over the expected background. The data from Gargamelle, at an average neutrino energy of 2 GeV, was later re-analyzed and presented at the NuInt 2002 conference\footnote{
To the best knowledge of the authors, details on the analysis methodology have been lost.}, leading to the only existing cross-section value for this channel ~\cite{Zeller:2003ey}.

Since the initial observation of these interactions decades ago, no new measurements have been reported. The absence of modern data, better aligned with the needs of neutrino oscillation experiments, limits our ability to validate neutrino interaction generators and assess the uncertainty associated to this channel.

In this article, we address this situation using data of the near detector ND280 of the T2K experiment. We report the strategy and selection algorithm performances used to identify the  largest existing sample of events for this channel, consisting of more than two hundred selected signal events. We present the observed event distributions in angle and momentum for the selected $\pi^+$ track, and compare them to simulation predictions. We describe the signal cross section extraction methodology and tests performed in the analysis to validate the robustness of the measurement and its statistical interpretation. Lastly, we report the measured double-differential signal cross section and compare it to T2K's simulation. Comparisons to alternative models are presented in Ref.~\cite{Abe:2025wxc}.

\section{Experimental Setup}
T2K~\cite{T2K:2011qtm} is a second generation accelerator-based long-baseline neutrino oscillation experiment in Japan, using a highly pure $\nu_\mu$ or $\bar{\nu}_\mu$ neutrino beam~\cite{T2K:2012bge} generated using facilities at the Japan Proton Accelerator Research Complex (J-PARC). The beam, described in more detail in Sec.~\ref{sec:beam}, is measured twice: once before and once after neutrino oscillations have occurred. Firstly, measurements are taken at a near detector facility at J-PARC located 280~m downstream from the neutrino production target. This facility consists of INGRID~\cite{Abe:2011xv} located on-axis, Wagasci-BabyMIND~\cite{Yasutome:2020ywt} at 1.5$^\circ$ off-axis, and ND280 at 2.5$^\circ$ off-axis, which is used in this study and described further in Sec.~\ref{sec:nd280}. Secondly, measurements are taken at Super-Kamiokande~\cite{Super-Kamiokande:2002weg}, a 50~kt water Cherenkov detector. Super-Kamiokande is located 295 km further downstream and placed at 2.5$^\circ$ off-axis. In this article, beam data measured by the ND280 detector is used.

\subsection{Neutrino Beam}
\label{sec:beam}
\begin{figure}[ht!]
\centering
\includegraphics[width=0.49\textwidth]{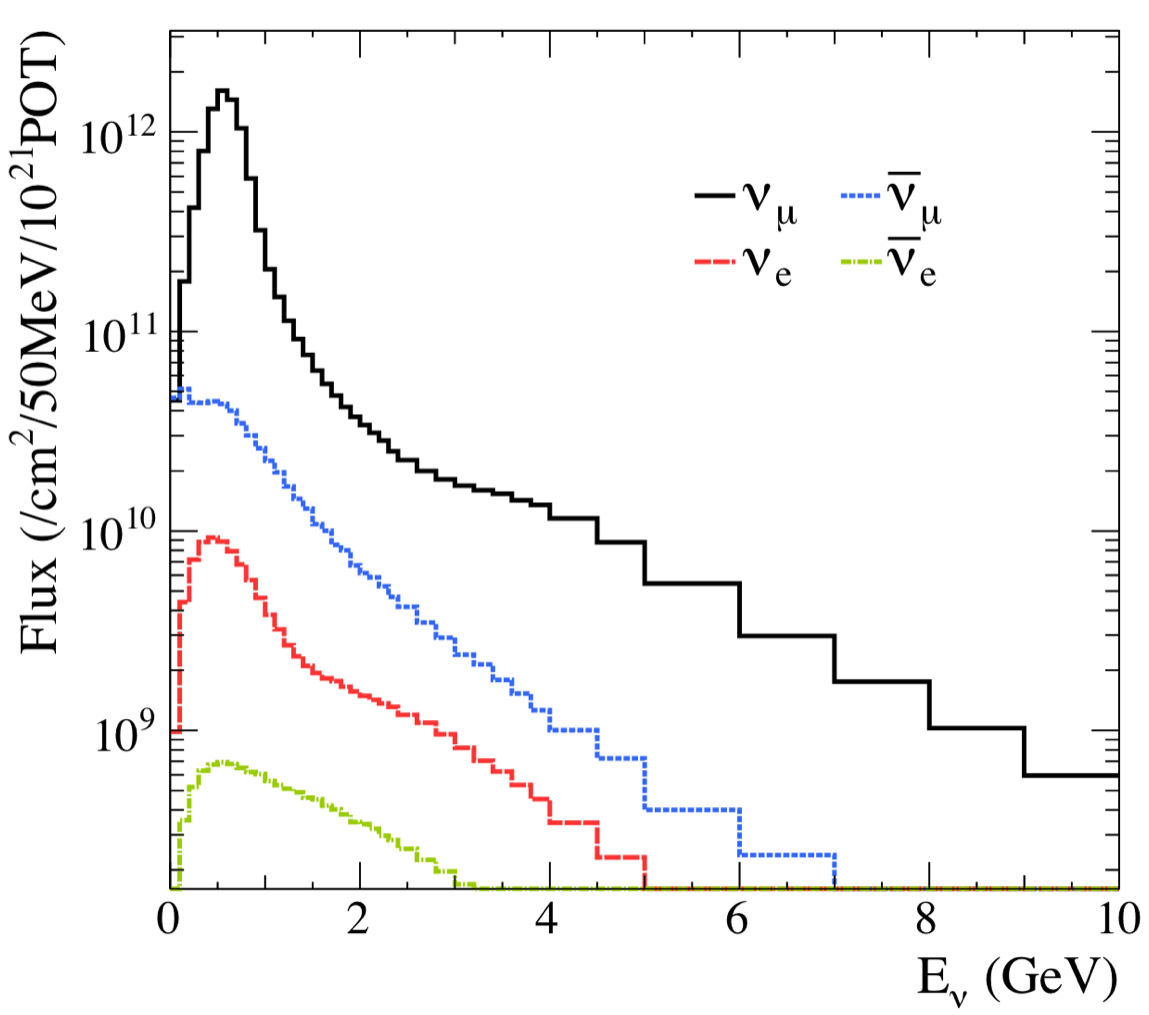}

\caption{T2K neutrino-mode flux prediction at the ND280 detector.}
\label{fig:flux}
\end{figure}
The T2K experiment uses a beam of neutrinos generated through the bombardment of a monolithic graphite target with a 30 GeV proton beam at the J-PARC facilities. The integrated data collection is measured in Protons on Target (POT), which quantifies the total exposure. The proton beam interactions with the target produce a secondary beam composed primarily of pions and kaons, which are focused using a set of three magnetic horns. T2K can run in either $\nu$ or $\bar{\nu}$ beam mode by focusing positive or negative mesons respectively. Mesons decay in a 96~m long helium-filled decay volume ending in a beam dump at the downstream end. High momentum muons crossing it are monitored using the MUMON detector~\cite{Suzuki:2012ova}. The meson decays result in the final beam of neutrinos, with a flux prediction at the ND280 detector site presented in Figure~\ref{fig:flux}. This study utilizes a total of $1.16\times10^{21}$~POT collected in $\nu$-beam mode. Because most neutrinos arise from pion decays via $\pi^\pm \rightarrow \mu^\pm + \nu (\bar{\nu})$, the neutrino energy spectra are strongly dependent on the beam off-axis angle~\cite{E899:1995bzq}. T2K exploits this feature by positioning ND280 and Super-Kamiokande, its main detectors for the oscillation analysis, at 2.5$^\circ$ off-axis, utilizing a narrow energy neutrino beam peaking at around 0.6 GeV. T2K's neutrino beamline has been recently upgraded~\cite{Igarashi:2021npv}; nevertheless, all of the data in this study corresponds to runs prior to the beam upgrade.

\subsection{The ND280 Neutrino Detector}
\label{sec:nd280}

\begin{figure}[ht!]
\centering
\includegraphics[width=0.49\textwidth]{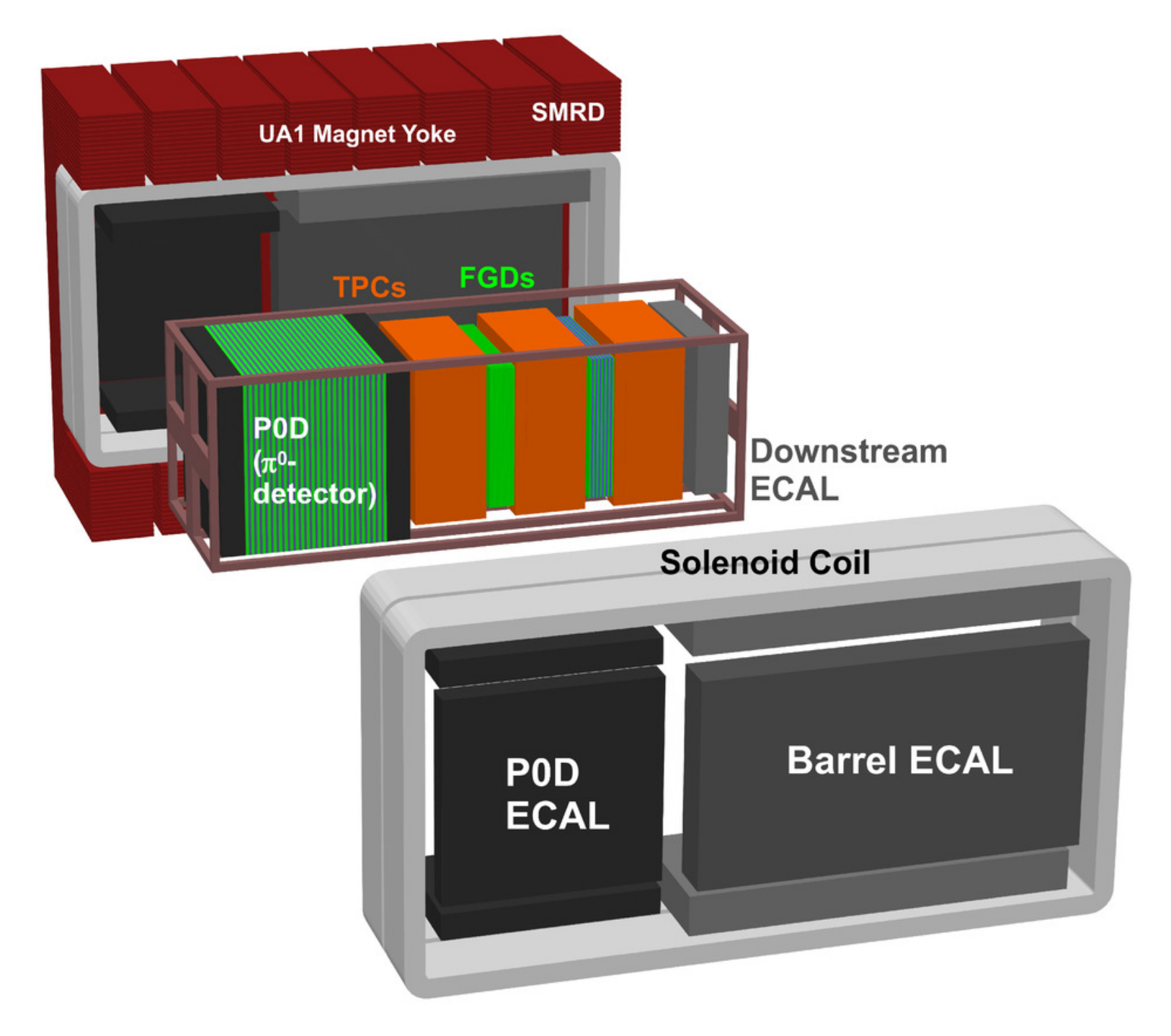}

\caption{Sketch of the ND280 detector in its original configuration, used to collect all the data in this study.}
\label{fig:nd280}
\end{figure}

The T2K Near Detector at 280 meters (ND280), presented in Figure~\ref{fig:nd280}, is a magnetized particle detector. It consists of sub-detector modules located inside the refurbished UA1/NOMAD magnet, providing a 0.2~T field used to bend charged particles moving inside ND280. The UA1 magnet yoke is instrumented with plastic scintillator slabs working as a Side Muon Range Detector (SMRD)~\cite{Aoki:2013swe}, and surrounding an electromagnetic calorimeter (ECal)~\cite{T2KUK:2013wkh} that nearly hermetically encapsulates its inner elements that are mounted on a steel frame. Inside it, going from upstream to downstream, the beam passes through the $\pi^0$ detector~\cite{Assylbekov:2011sh}, and the inner ND280 tracker consisting of three gaseous Time Projection Chambers (TPC1, 2, 3)~\cite{T2KND280TPC:2010nnd}, interleaved with two Fine-Grained Detectors (FGD1, 2)~\cite{T2KND280FGD:2012umz}. FGD1 is a plastic scintillator detector, consisting of 15 planes of two layers. Every layer is made up of 192 parallel bars, with consecutive layers using bars with alternate orientations along the two axes perpendicular to the neutrino beam. FGD2 follows a similar layout to that of FGD1, but alternating seven plastic scintillator planes with six inactive water bags.\\
The FGDs provide a total active target mass close to two tons, while the TPCs are used to identify different types of particles and their kinematics by measuring their curvature and $dE/dx$. In the measurement reported in this article, interactions starting in the FGD1 are selected. To increase the signal purity, the fiducial volume of FGD1 excludes 6 bars at each side of the detector as well as the first and the last planes.

\section{Event Simulation and Signal Selection}

Proton beam interactions with the graphite target and the propagation of daughter particles through the target station are modeled using FLUKA~\cite{Ferrari:2005zk, Bohlen:2014buj}. Subsequent interactions and decays in the beamline are modeled with GEANT3~\cite{Brun:1994aa} and GCALOR~\cite{Zeitnitz:1992vw} to predict the neutrino flux. The run-by-run flux simulation is tuned to the beam conditions as recorded by the beam monitoring systems and hadronic interactions and multiplicities are tuned to the NA61/SHINE hadron production measurements with a T2K replica target~\cite{NA61SHINE:2016nlf, NA61SHINE:2018rhe}. Neutrino interactions in the ND280 detector including the description of all the final-state particle kinematics are simulated using the NEUT~v5.4.0~\cite{Hayato:2009zz, Hayato:2021heg} and GENIE~v2.8.0~\cite{Andreopoulos:2009rq} neutrino event generators, hereafter NEUT and GENIE, using respectively about 20 (10) times more POT than the data. An in-depth description on the details for the NEUT (GENIE) model can be found in Ref.~\cite{T2K:2023smv} (Ref.~\cite{T2K:2020lrr}). The propagation of the final-state particles through the detector medium is simulated using GEANT4~v4.9.4~\cite{GEANT4:2002zbu, Allison:2006ve, Allison:2016lfl} with the \texttt{QGSP\_BERT} physics list~\cite{Allison:2016lfl}. Pion secondary interactions are handled by the cascade model in NEUT and treated as a detector systematic uncertainty.

\subsection{Signal and Background Definitions}
Simulated events are classified according to the following definitions, based on their true information after Final-State Interactions (FSI), i.e.\ considering only particles that exit the nucleus. The categories are built based on the particle content irrespectively of its kinematic except for the NC1$\pi^+$ 0p and NC1$\pi^+$ Np categories are descried below:
\begin{itemize}
\setlength\itemsep{-0.em}
\item \textbf{$\nu_\mu$ CC0$\pi$:} All $\nu_\mu$ interactions with a muon and without mesons in the final state.
\item \textbf{$\nu_\mu$ CC1$\pi^+$:} All $\nu_\mu$ interactions with a muon and a positive pion and no other mesons in the final state.
\item \textbf{$\nu_\mu$ CC-other:} All $\nu_\mu$ interactions with a muon which are not included in other topologies.
\item \textbf{$\bar{\nu}_\mu$ CC:} All $\bar{\nu}_\mu$ interactions with an antimuon in the final state.
\item \textbf{$\nu_e$/$\bar{\nu}_e$ CC:} All $\nu_e$ and $\bar{\nu}_e$ interactions with an electron or positron in the final state.
\item \textbf{NC0$\pi$:} All $\nu$  and $\bar{\nu}$ interactions without charged leptons nor mesons in the final state.
\item \textbf{\textcolor{black}{NC1$\pi^+$ 0p (Signal):}} All $\nu$ and $\bar{\nu}$ interactions without charged leptons, a single positive pion, no protons with true momentum $>200$~MeV/$c$ and no other mesons nor other charged particles in the final state. Any number of neutrons is allowed. The cut on true proton momentum is discussed in the following section (Sec.~\ref{sec:pr_threshold}).
\item \textbf{NC1$\pi^+$ Np:} All events that would be signal (NC1$\pi^+$ 0p) but don't satisfy the true proton momentum condition.
\item \textbf{NCX$\pi^0$:} All $\nu$ and $\bar{\nu}$ interactions without charged leptons and at least one neutral pion in the final state. Any number of other mesons are allowed.
\item \textbf{NC-other:} All $\nu$ and $\bar{\nu}$ interactions without a charged lepton that are not included in any other topology.  
\end{itemize}

\subsection{Signal Selection Method}
\begin{figure}[ht!]
\centering
\includegraphics[width=0.49\textwidth]{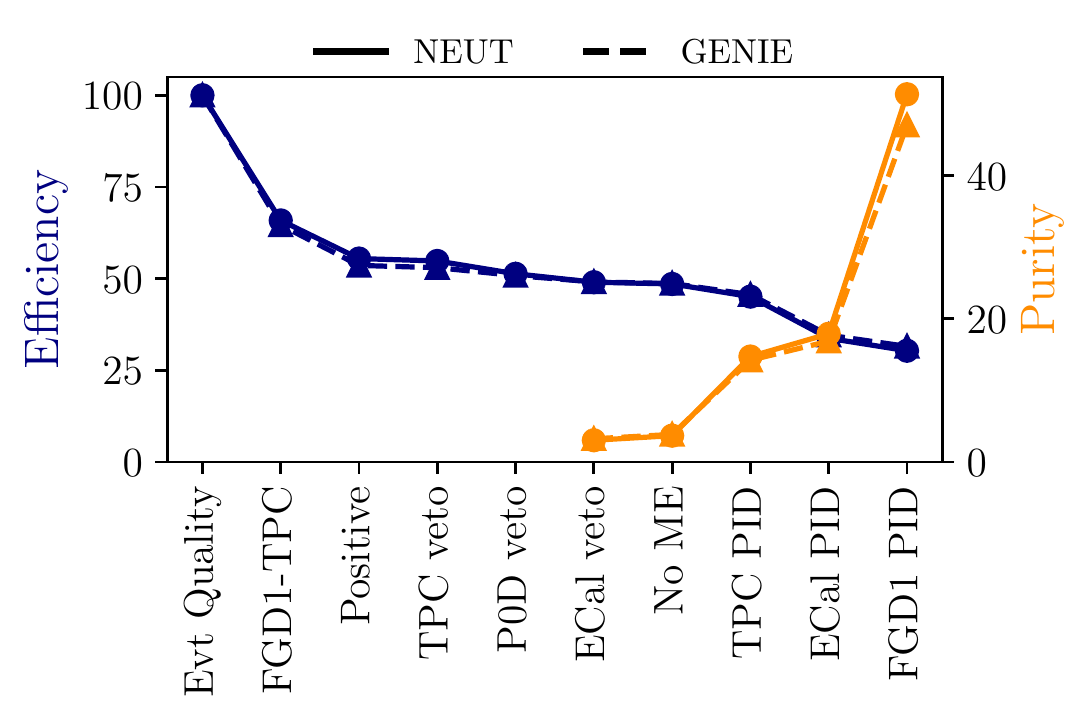}
\caption{Sequential evolution of the efficiency and purity of the selection for the signal sample.}

\label{fig:redPS_eff_pur_vs_cut}
\end{figure}

Due to its low cross section, NC1$\pi^+$ events are expected to make up only 1.3\% of all interactions in FGD1. A selection algorithm was developed to identify a signal sample enriched in signal events, complemented by background-enriched samples. The logic of the selection algorithm, associated with Figure~\ref{fig:redPS_eff_pur_vs_cut}, is as follows:
\begin{enumerate}
    \item \textbf{FGD1-TPC:} The main track in the event, i.e.\ the only track that exits FGD1, must reach at least one ND280 TPC, creating a reconstructed track segment. This TPC segment must have at least 19 hit clusters, a standard criteria in analyses, to ensure reliable TPC reconstructed information. Events that consist exclusively of FGD1-contained tracks are rejected. Events with multiple tracks that escape the FGD1 volume are rejected, as for signal events only one track is expected. Events with one additional track contained in FGD1 are retained separately, in a dedicated sample, that will be referred to as the Additional Track (\texttt{AddTrk}) sample.
    
    \item \textbf{Positive:} The track trajectory in the TPC must show curvature matching a forward-moving, positively charged particle according to ND280's magnetic field.

    \item \textbf{Vetos:} To reduce background from events outside the FGD1 target, called out-of-fiducial volume (OOFV) events, any events with tracks reconstructed upstream of FGD1 are rejected. Three sequential vetos are applied based on the presence of reconstructed tracks in the most upstream TPC, the pi-zero detector, and the upstream part of the ECal.
 
    \item \textbf{No ME:} No Michel Electron (ME) signatures are identified in FGD1. ME are produced in $\mu$ decays and are identified looking for delayed hits. For FGD1, their detection efficiency is $64.2\pm2.0$~\%.
    \item  \textbf{TPC PID:} The TPC particle identification (PID) information ---based on the measured $dE/dx$ with respect to the particle momentum measured by curvature--- must be compatible with that of a Minimum Ionizing Particle (MIP). Tracks rejected by this cut are retained separately in the TPC PID (\texttt{TPID}) sample. 
    \item \textbf{ECal PID:} For those tracks reaching the ECal, their PID information must be compatible with a hadron. The ECal offers excellent capabilities in separating $\mu$ vs charged $\pi$, and in consequence, rejected events are collected in a separate control sample, named the ECal PID (\texttt{EPID}) sample. 
    \item \textbf{FGD1 PID:} Lastly, the main track $dE/dx$ in FGD1 must be compatible with a MIP. This cut disentangles the cases where a $\pi^+$ exits FGD1 and the cases where a negatively charged particle travels backwards and stops in the FGD1.
\end{enumerate}

Figure~\ref{fig:redPS_eff_pur_vs_cut} shows the efficiency and purity of the event selection for events with $\cos\theta_{\pi^{+}}>$0.5 and $0.2<p_{\pi^+}< 1.0$~GeV/$c$ (see Sec.\ref{sec:kin_considerations}). The main efficiency drops correspond to the necessary requirement of having the candidate $\pi^+$ reaching the TPC and to the ECal PID cut. Notably, events rejected by the ECal PID are largely retained in the \texttt{EPID} sample. The purity, negligible before applying PID cuts, increases drastically with the last cuts in the selection, demonstrating an effective enhancement of the fraction of signal events with a minor decrease in efficiency. The impact of the cuts on the signal efficiency and purity are well-matched in NEUT and GENIE simulated events demonstrating that the selection process is robust to the simulation physics model.

\subsection{Proton Threshold for Signal Events}
\label{sec:pr_threshold}
\begin{figure}[ht!]
\centering
\includegraphics[width=0.49\textwidth]{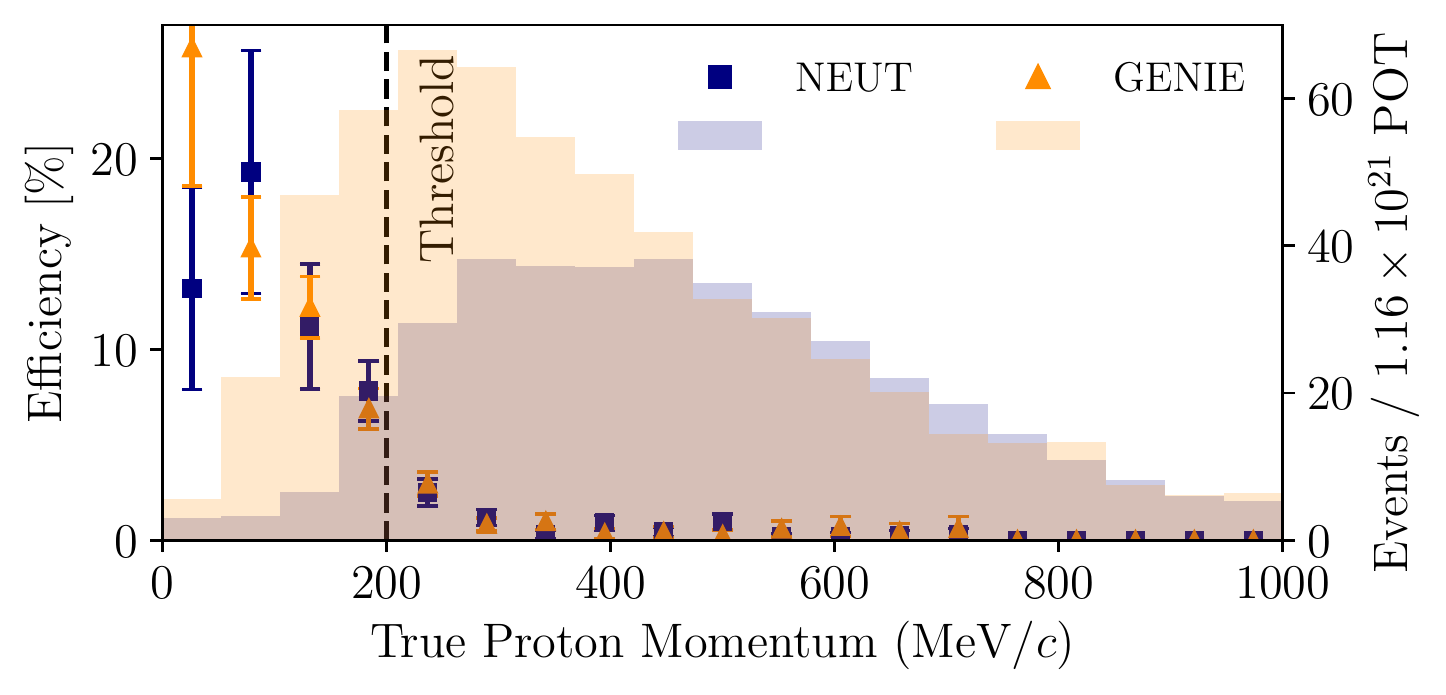}
\caption{Selection efficiency for NC1$\pi^+$ events with protons (markers, left Y-axis) as a function of true proton momentum. Error bars indicate statistical uncertainty.} The momentum distribution of the highest-momentum proton in NC1$\pi^+$ events prior to any selection is shown as a shaded histogram (right Y-axis).

\label{fig:proton_definition}
\end{figure}

Low momentum protons, arising from nuclear effects including FSI, are often merged during the reconstruction to the other reconstructed tracks modifying their observed light yield. In consequence, the FGD1 PID cut, which depends on the observed light yield, is sensitive to low momentum protons below the tracking threshold. The efficiency to select NC1$\pi^+$ events with protons is presented in Fig.~\ref{fig:proton_definition}. 
NC1$\pi^+$ events with proton momentum below 200 MeV/$c$ are sometimes selected and are therefore included in the signal definition to be robust against model prediction discrepancies. Events with protons above 200 MeV/$c$ are rarely selected, with much lower selection efficiency than other NC1$\pi^+$ events. Therefore, they are treated as background. NEUT, modeling FSI using a semi-classical intra-nuclear cascade model, and GENIE using the ``hA'' model, render noticeably differences as observed Fig.~\ref{fig:proton_definition}. To validate the proton threshold choice, fake data studies modifying the number of background NC1$\pi^+$ events with protons were performed. As later reported in Sec.~\ref{sec:model_indep}, no measurement bias was observed.

\subsection{Kinematic Considerations About The Positive Pion}
\label{sec:kin_considerations}
\begin{figure}[ht!]
\centering
\includegraphics[width=0.49\textwidth]{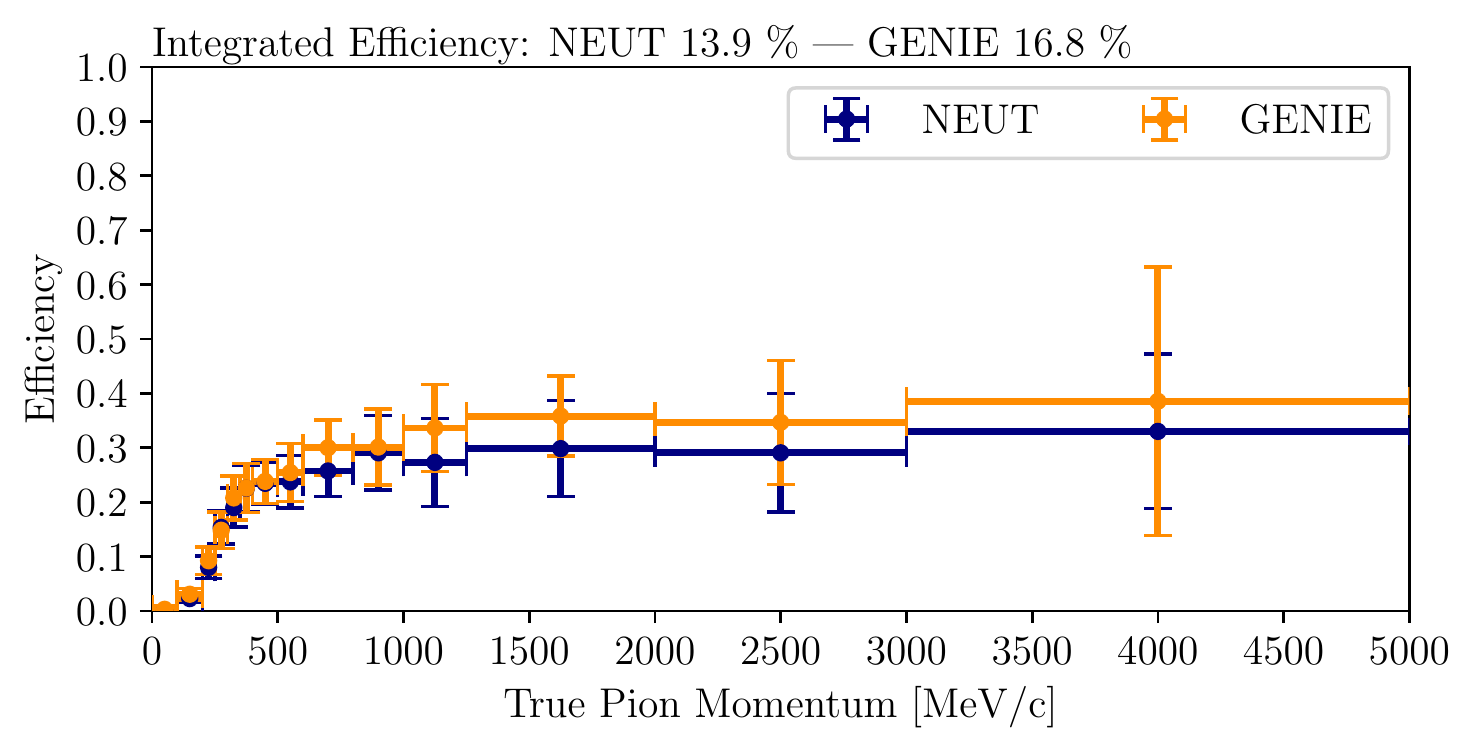}
\includegraphics[width=0.49\textwidth]{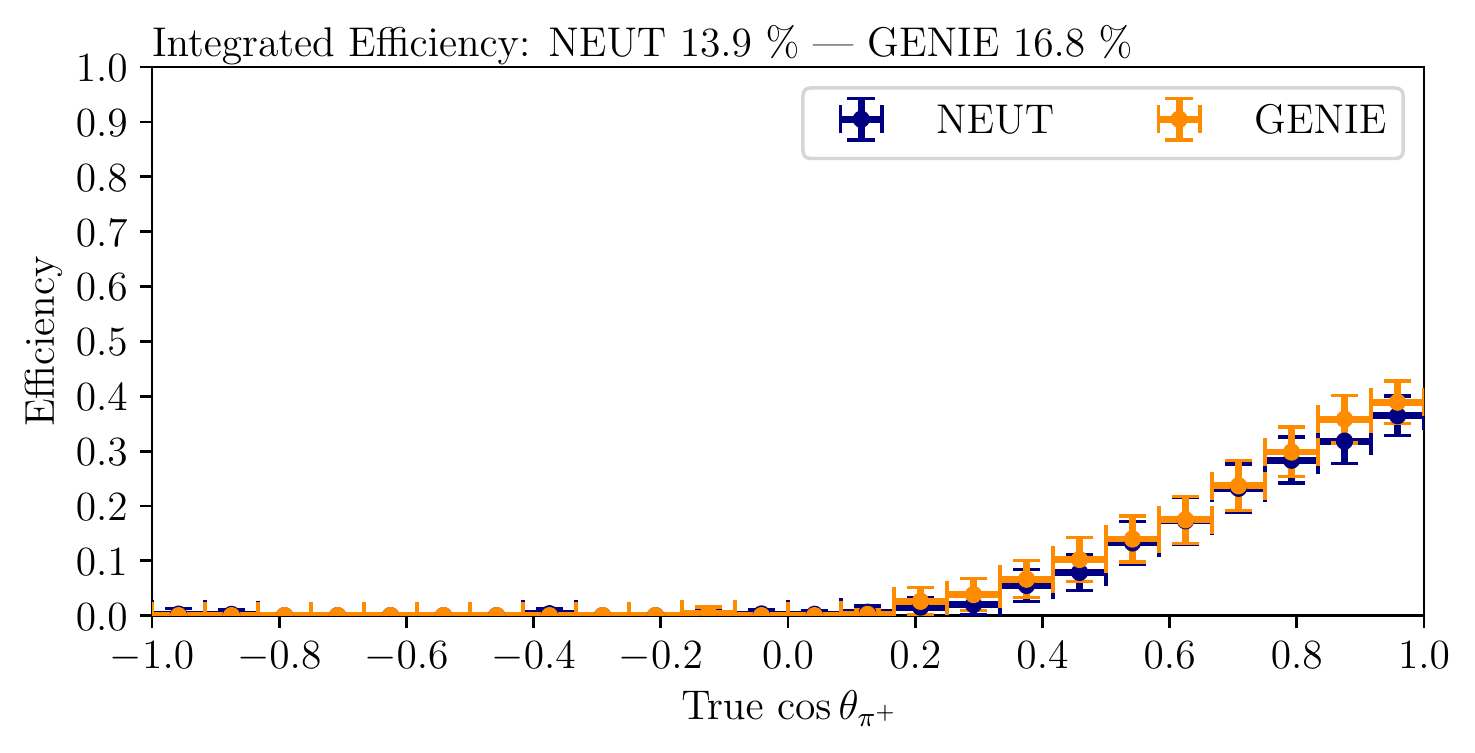}
\includegraphics[width=0.49\textwidth]{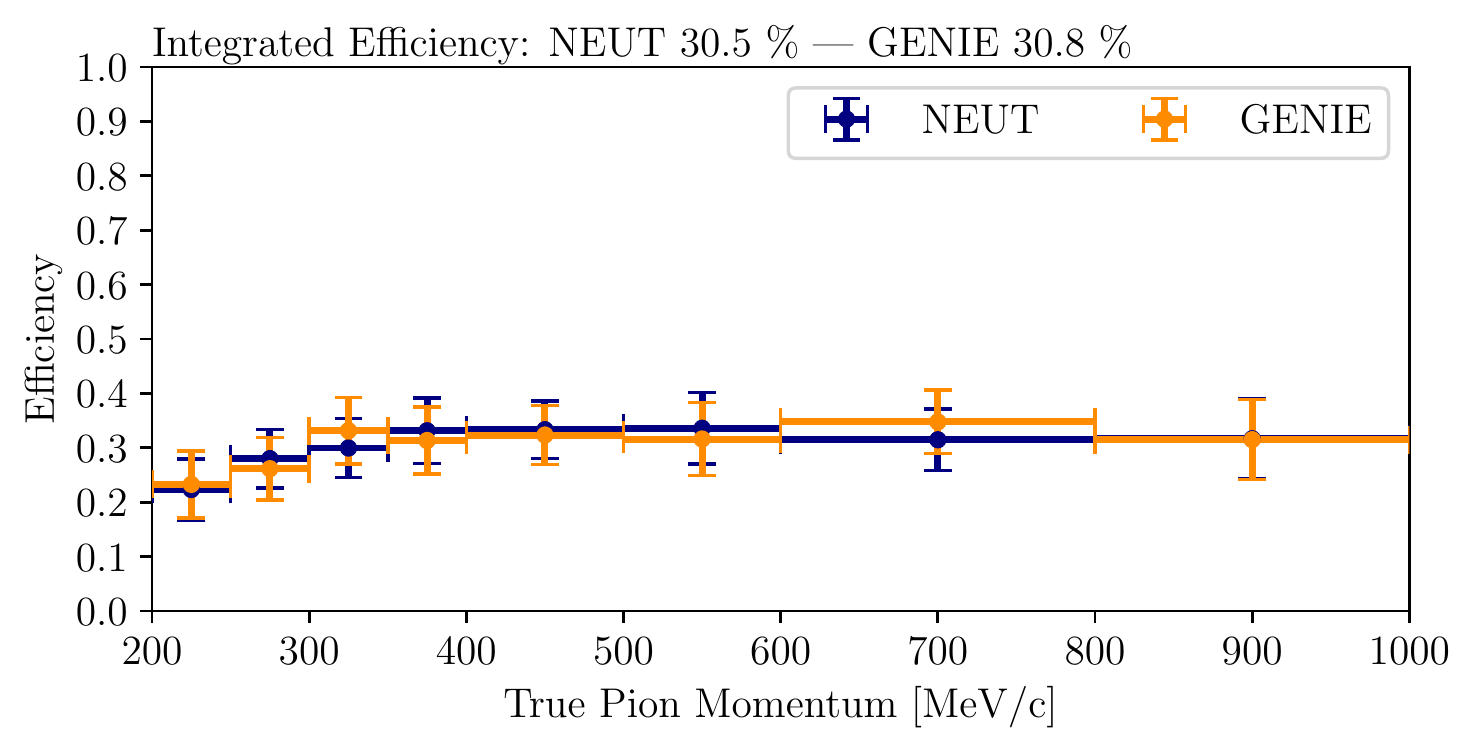}
\includegraphics[width=0.49\textwidth]{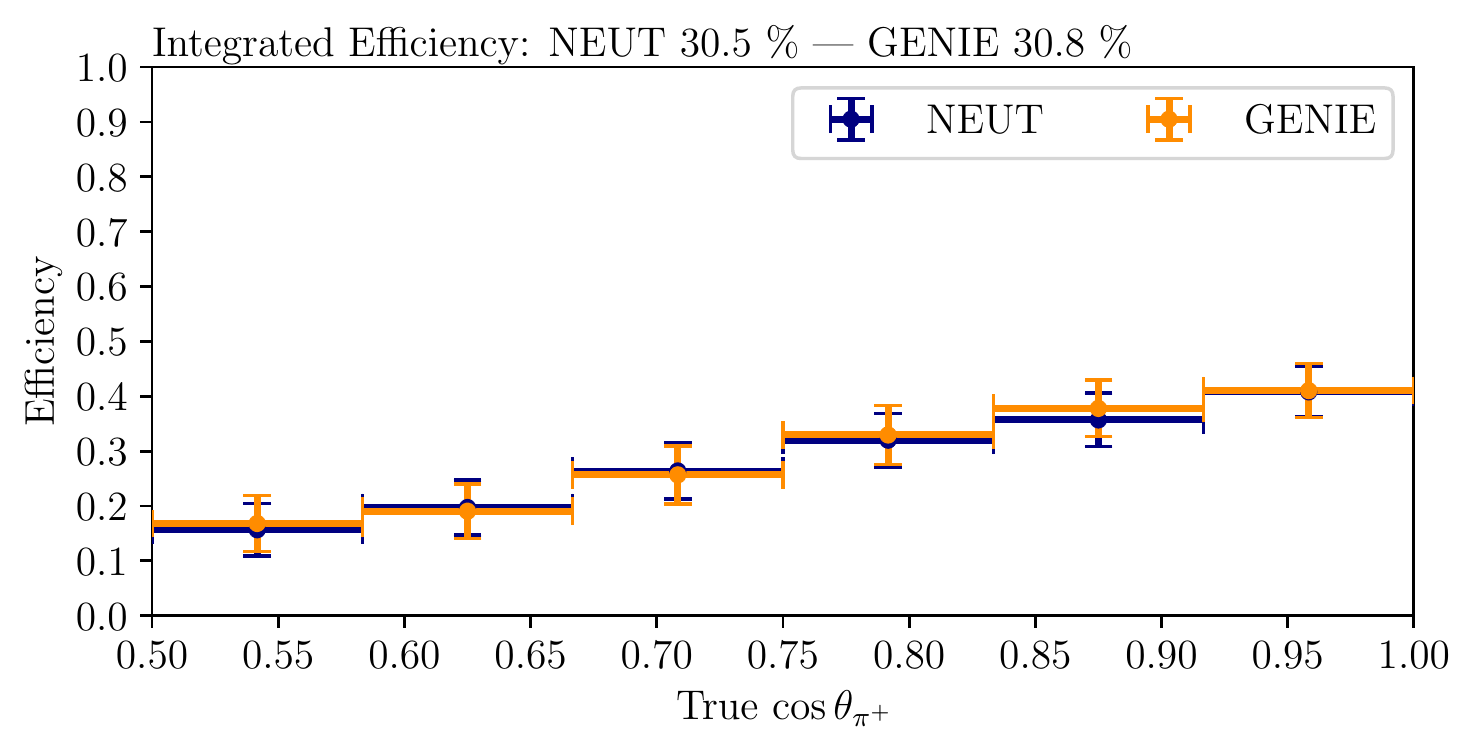}
\caption{Projected 1D efficiencies as a function of the true $\pi^+$ momentum and angle. The top two panels are for the full phase space and the bottom two for phase space restricted to the RoI, see the text for details. Error bars show the MC statistical uncertainty.}
\label{fig:effs}
\end{figure}

The selection algorithm requires that all events have TPC information for the main track. A veto is applied to those events with activity in the most upstream TPC, and accordingly, only NC1$\pi^+$ events with a forward-going $\pi^+$ can be selected. The selection efficiency is highest for small angles with respect to the beam, as the probability of reaching TPC2 decreases with increasing angle.
The probability of a track entering the TPC2 from FGD1 also increases with momentum. However, even in the most favorable case, when the $\pi^+$ is forward-going and the interaction vertex is in the downstream region of FGD1, $\pi^+$ tracks below 200 MeV/$c$ are unlikely to be successfully reconstructed in the TPC. These effects can be clearly observed in the selection efficiency, presented in Fig.~\ref{fig:effs}. NC1$\pi^+$ events have three particles in the final state, whereas most of the competing backgrounds have only two. Therefore, for the same neutrino energy, the main track in NC1$\pi^+$ is expected to have lower momentum than that of most background events and consequently the purity is best for low track momenta. Accordingly, we focus on studying and measuring the NC1$\pi^+$ cross section defined by $\cos\theta_{\pi^{+}}>$0.5 and $0.2<p_{\pi^+}< 1.0$~GeV/$c$, which will hereafter be referred to as the Region of Interest (RoI).

Looking again to Fig.~\ref{fig:effs}, the integrated selection efficiency, unrestricted by any pion kinematics, is 13.9\% for NEUT and 16.8\% for GENIE. The discrepancies between NEUT and GENIE are mostly observed outside the RoI, where the selection behavior is very similar for both generators. The integrated efficiency for events within the RoI is of 30.5\% for NEUT and 30.9\% for GENIE.

\subsection{Signal Sample}

\begin{figure}[ht!]
\centering
\includegraphics[width=0.49\textwidth]
{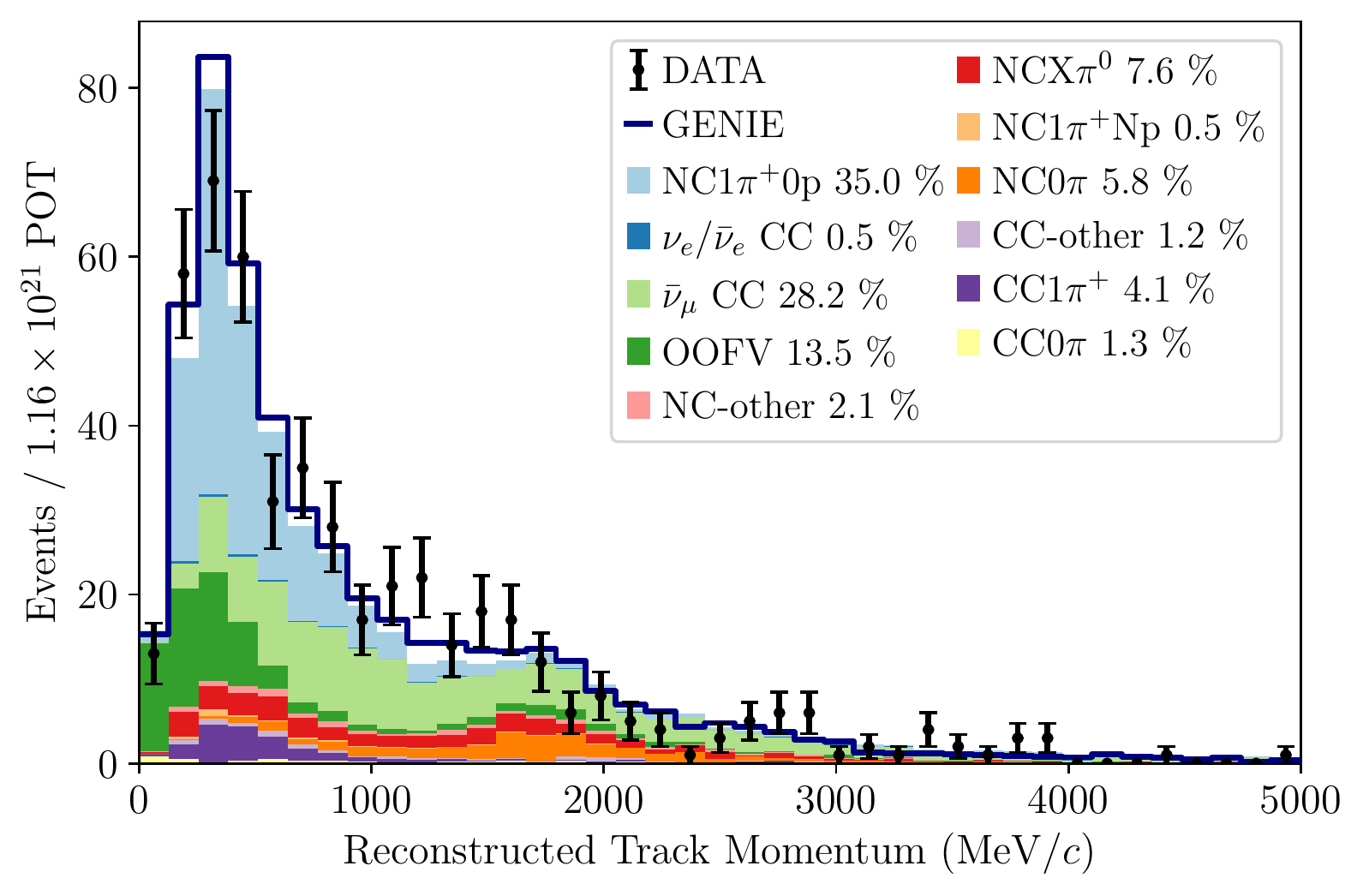}
\hfill
\includegraphics[width=0.49\textwidth]{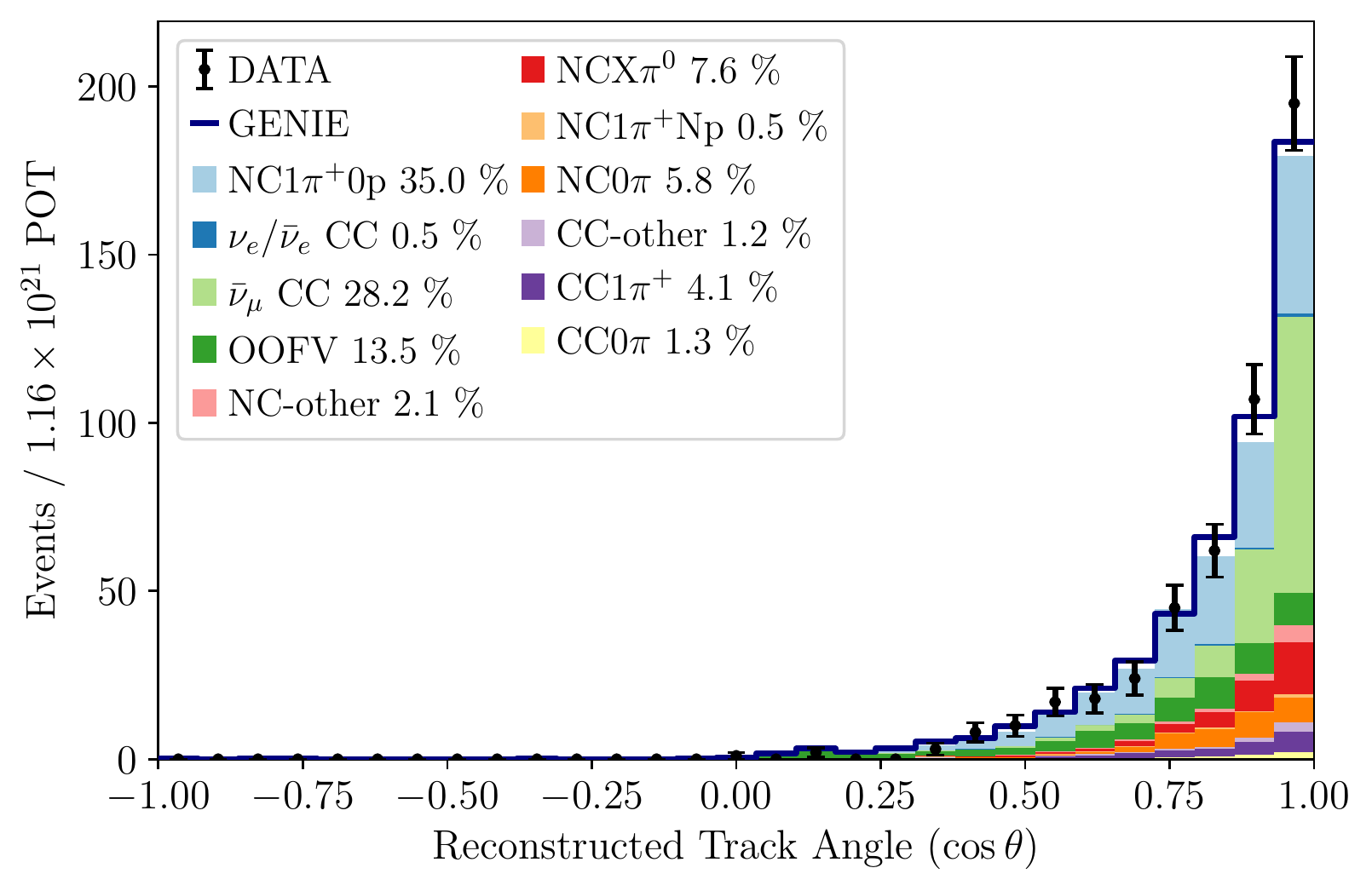}

\caption{One dimensional pion momentum and angle distributions for the signal-enriched sample. NEUT predictions are presented as a stacked histogram for the different event topologies, and are compared in every bin to the total event rate expected according to GENIE and observed in T2K data.}
\label{fig:signal}
\end{figure}

The selected signal sample is presented in Figure~\ref{fig:signal}. The combination of the selection cuts effectively achieves the goal of selecting a sample of data enriched in NC1$\pi^+$ interactions, with an integrated purity in the RoI of 51.4\% (47.1\%) for NEUT (GENIE), which reaches about 60\% purity at  the peak of the pion momentum distribution around 300 MeV/$c$. The primary feature of the selected events consists for both NEUT and GENIE of a strong signal peak in the low momentum region outstanding above a relatively flat background component. This structure can be clearly observed in data. The main background for the selection consists of $\bar{\nu}_\mu$ CC events, typically consisting of a $\mu^+$ misidentified as a $\pi^+$. Despite the small fraction of $\bar{\nu}$ in the neutrino beamline (2.3\%) when T2K operates in $\nu$-mode, $\bar{\nu}_\mu$ CC events before any selection cuts are twice as numerous as the signal events. At low reconstructed momentum, the dominant background are FGD1 out-of-fiducial volume events that consist of an aggregate of multiple processes: some arise from vertex migration failures during reconstruction (27.7~\%), with the rest occurring via secondary interactions of $\pi^\pm$ (17.5~\%), neutron (45.2~\%) and other particles (9.9~\%). Also in that region a small population of CC1$\pi$ events is present, expected for those events where a low momentum muon is not reconstructed and its Michel electron signature is not detected. A small population of NC0$\pi$ events is present around 1.8 GeV/c, where proton tracks are MIP-like, making them indistinguishable from charged pions in the TPC. Accordingly, all selected particle populations are understood and are the direct result of the strengths and limitations of the ND280 detector.

Table~\ref{tab:events_in_signal_sample_by_topo} shows the total expected events for NEUT and GENIE. The expected signal events double the largest previous existing sample, provided by the Gargamelle~\cite{GargamelleNeutrinoPropane:1977hya}. An additional 56.3 (53.1) signal events are selected in the \texttt{EPID} sample according to NEUT (GENIE), with an overall purity similar to 10\%, reaching 30\% at around 300 MeV/$c$.  In total, this analysis selects over 200 NC$1\pi^+$ events, most of them in regions with significant purity.

\begin{table}[hbt!]
    \footnotesize
    \centering
    \caption{Number of expected selected events according to NEUT and GENIE scaled to the data POT. Signal events are highlighted in bold.}
\begin{tabular}{lrr}
\hline \hline
  		 &   NEUT    & GENIE   \\
Topology &   Events  & Events  \\
\hline \hline
CC0$\pi$ &                6.3 &      9.6 \\    
CC1$\pi^+$ &              19.4 &     24.9 \\   
CC-other &                5.8 &      3.7 \\    
NC0$\pi$ &                27.4 &     20.2 \\   
\textbf{NC1$\pi^+$0p} &            164.3 &  165.9 \\
NC1$\pi^+$Np &            2.5 &      4.3 \\    
NCX$\pi^0$ &              35.7 &     44.5 \\   
NC-other &                10.0 &      6.8 \\    
OOFV &                    63.3 &     74.7 \\   
$\bar{\nu}_{\mu}$ CC &    132.5 &  133.9 \\
$\nu_e/\bar{\nu}_e$ CC &  2.3 &      2.3 \\
\hline \hline
\end{tabular}
    \label{tab:events_in_signal_sample_by_topo}
\end{table}

\subsection{Background Samples}

\begin{figure*}[ht!]
\centering
\includegraphics[width=0.49\textwidth]{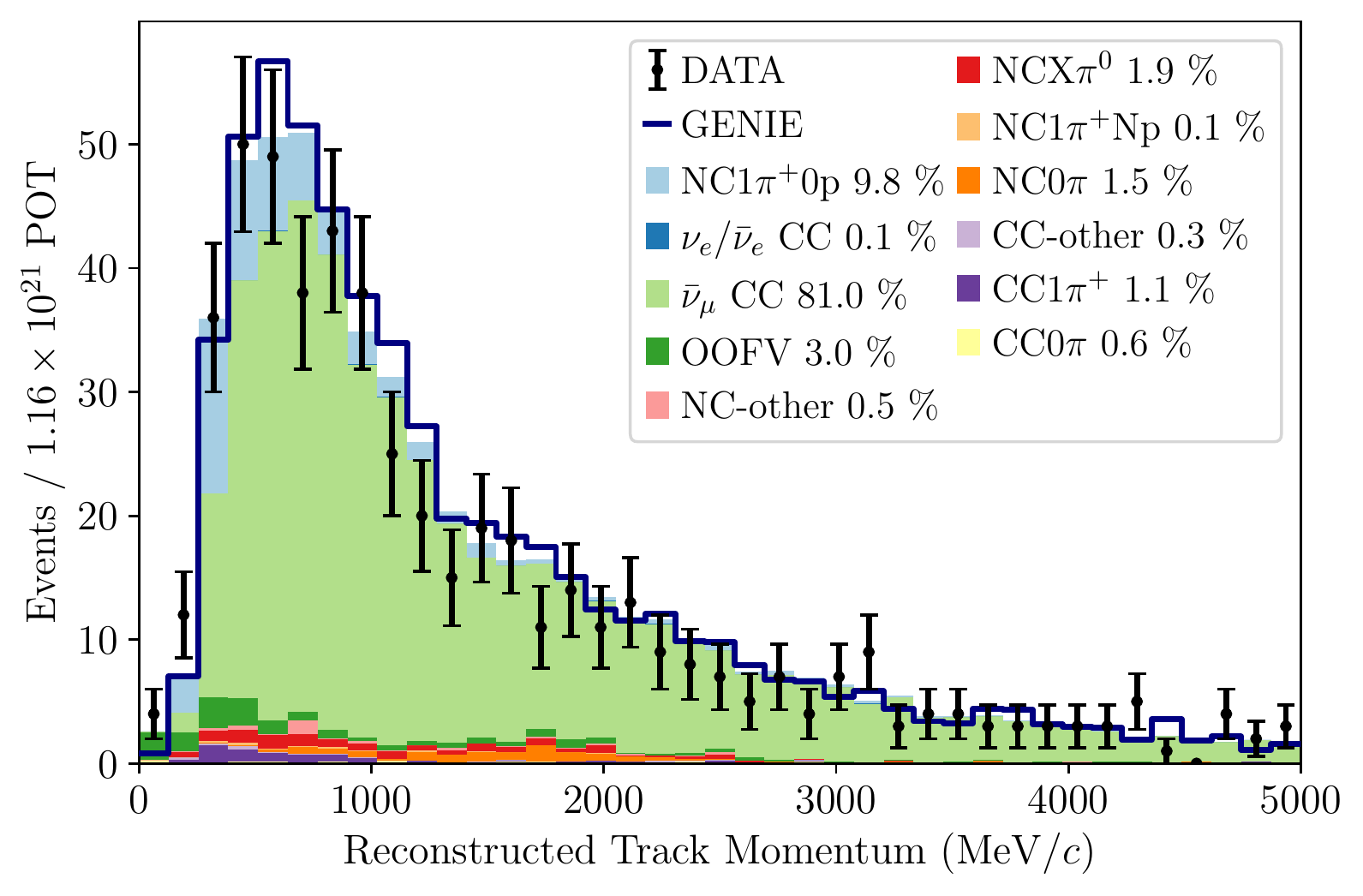}
\hfill
\includegraphics[width=0.49\textwidth]{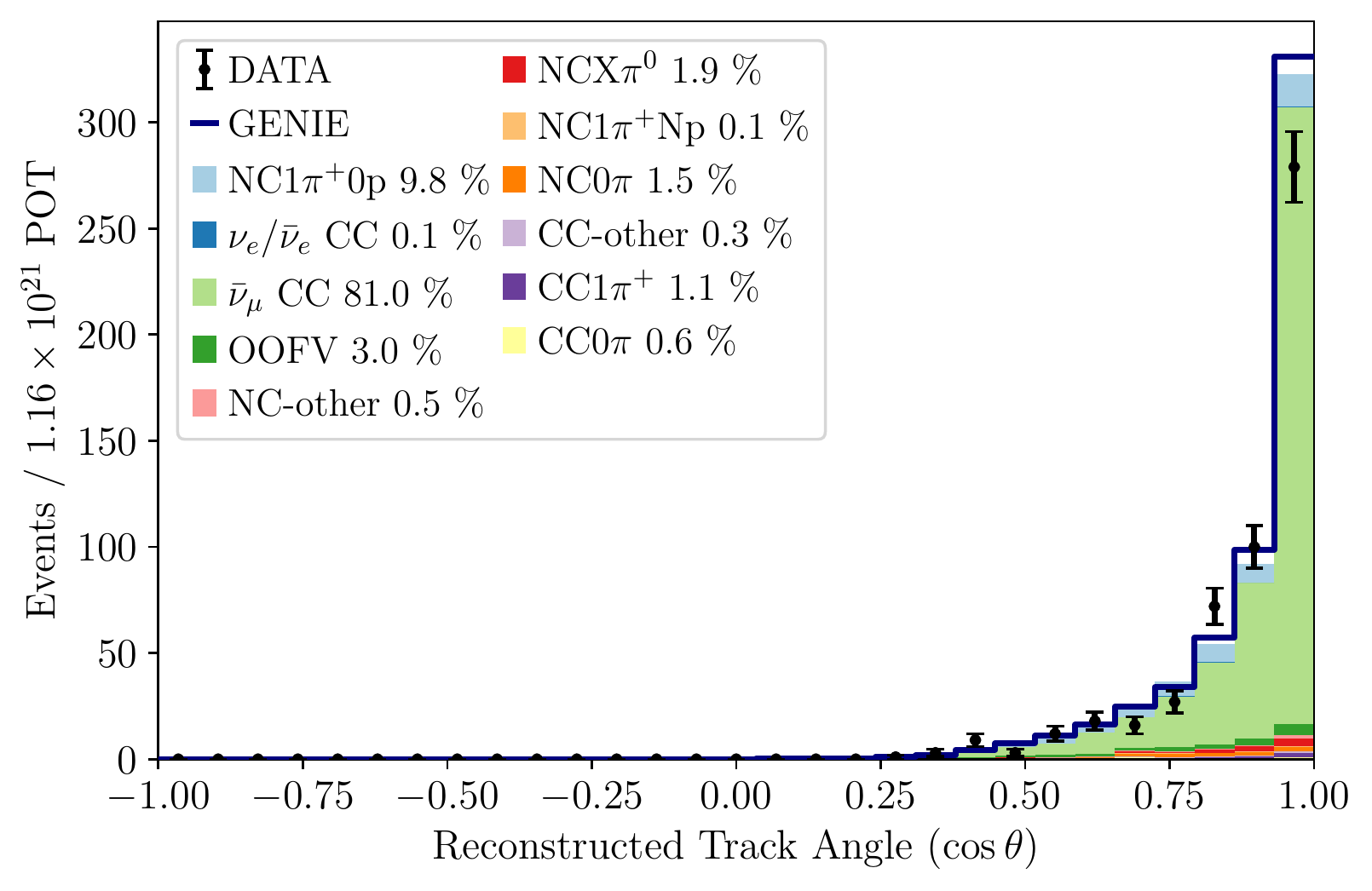}

\includegraphics[width=0.49\textwidth]{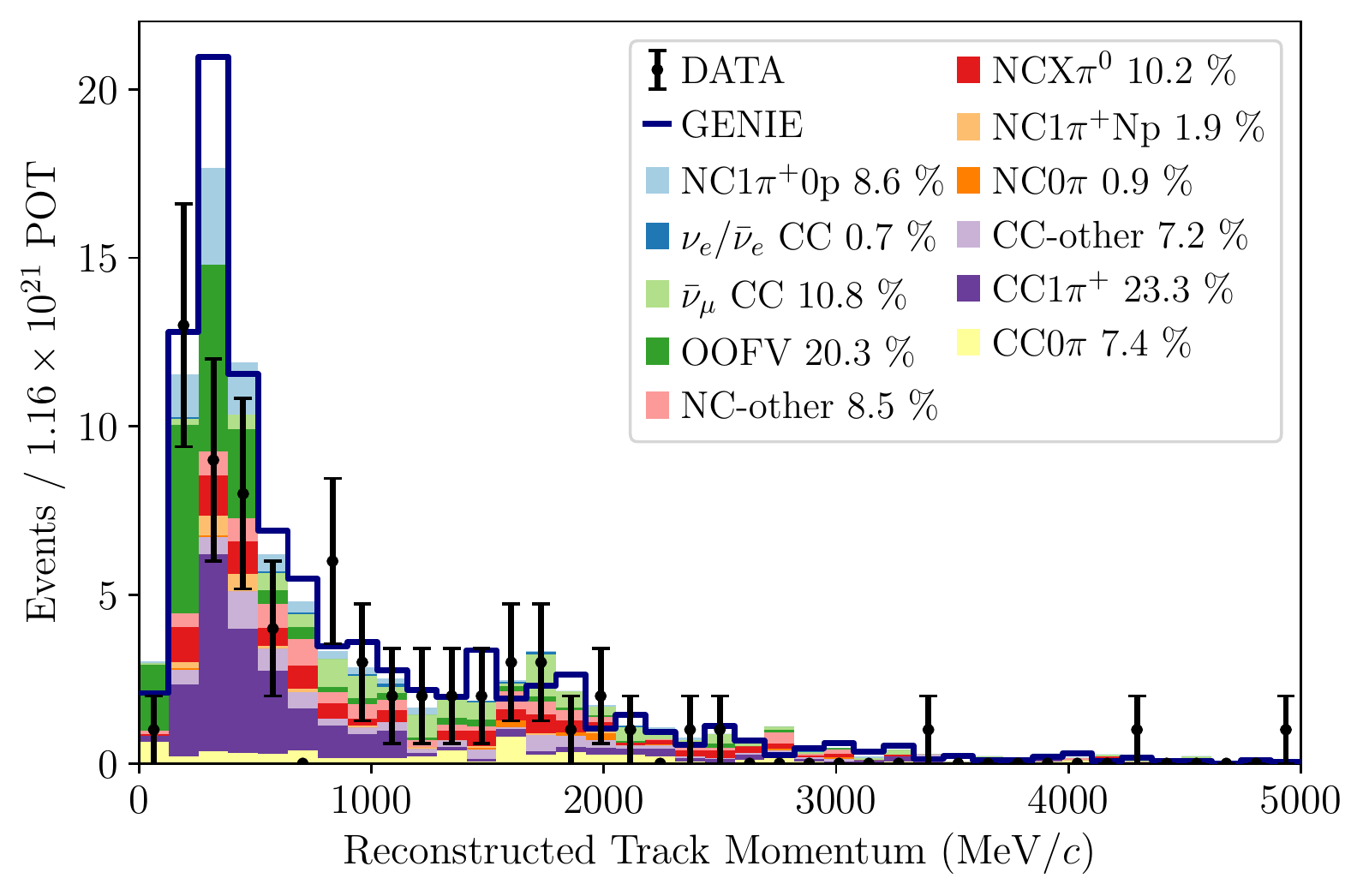}
\hfill
\includegraphics[width=0.49\textwidth]{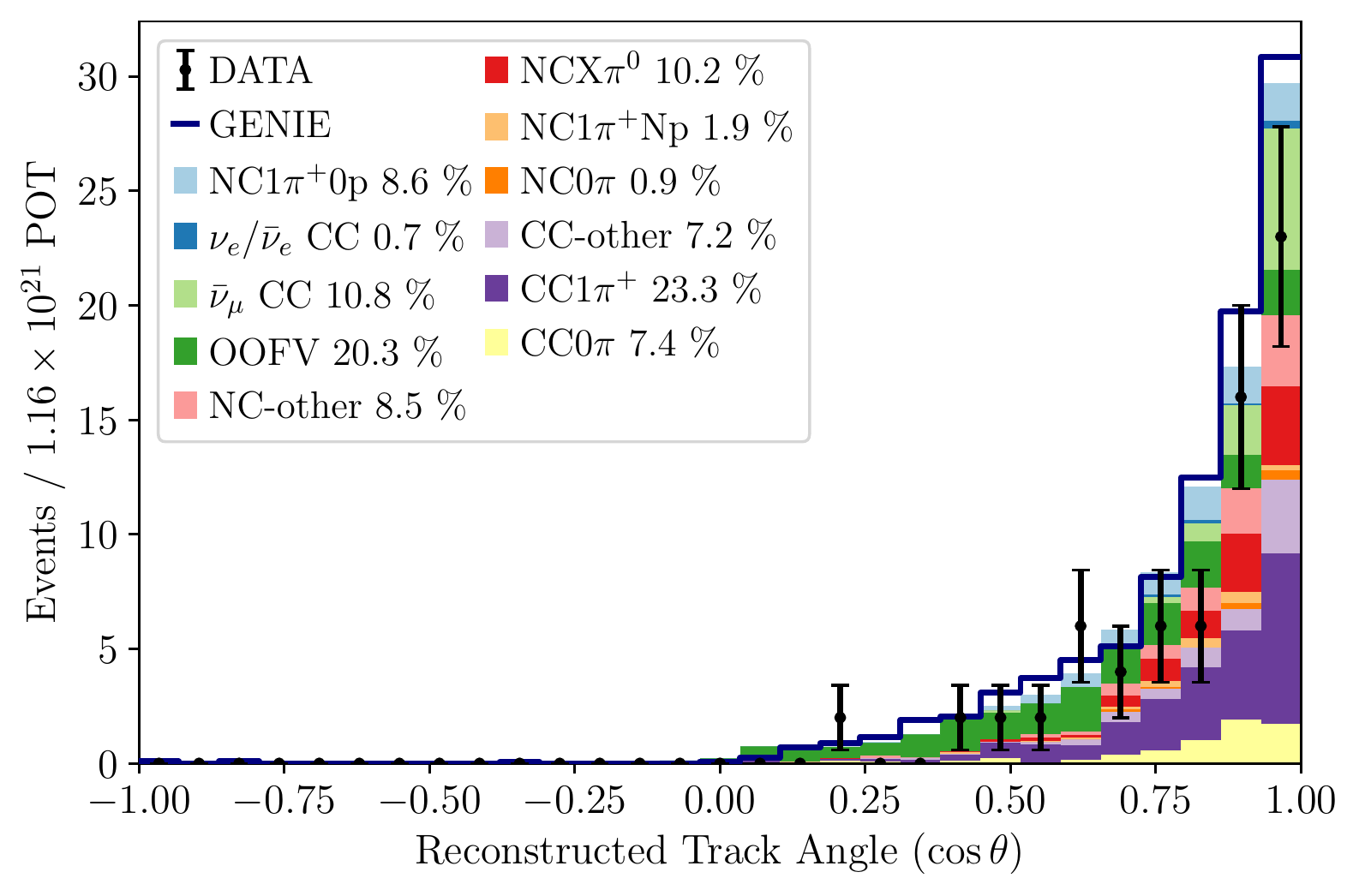}

\includegraphics[width=0.49\textwidth]{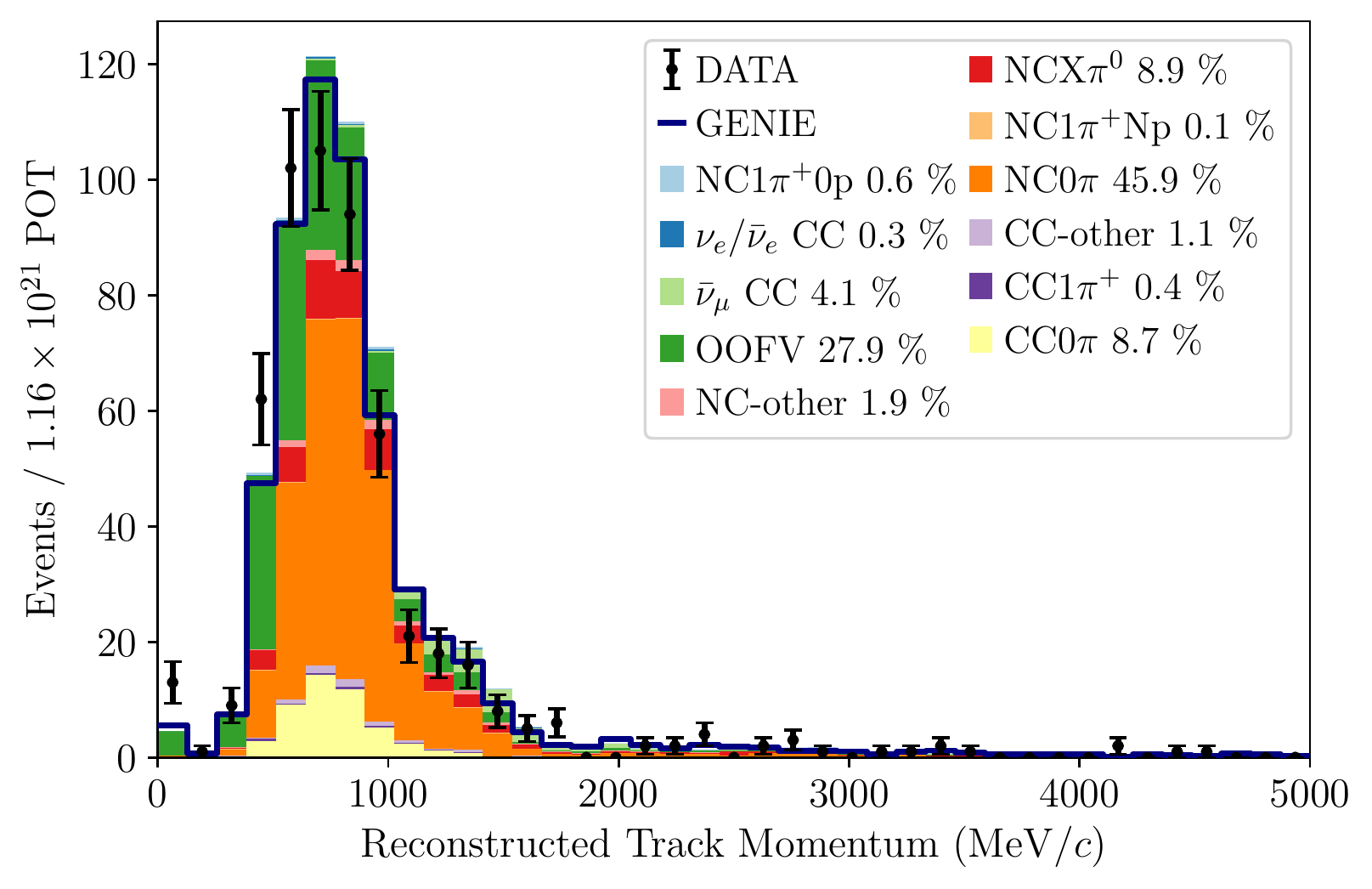}
\hfill
\includegraphics[width=0.49\textwidth]{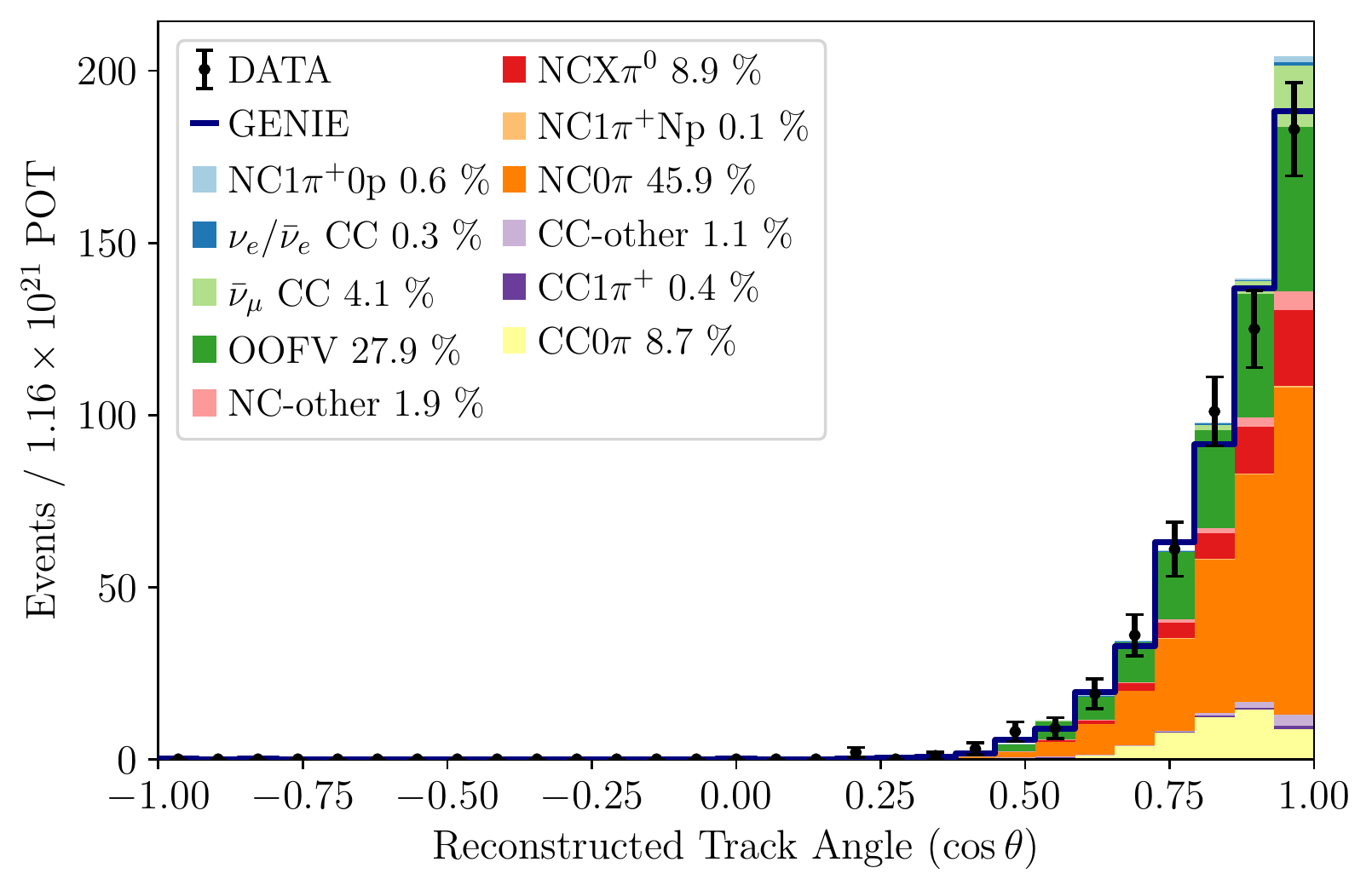}

\caption{One dimensional momentum and angle distributions for the background-enriched sample. From top to bottom the plots in each row corresponds to the \texttt{EPID}, the \texttt{AddTrk} and the \texttt{TPID} sample.}

\label{fig:control_regions}
\end{figure*}

The expected event rates according to NEUT and GENIE for the various background samples are compared to data in Figure~\ref{fig:control_regions}. Background samples offer a handle on important background events, particularly CC1$\pi$, $\bar{\nu}$~CC and NC0$\pi$ events. The main characteristics of these samples are:
\begin{itemize}
    \item \texttt{EPID}: Consists of signal-like events where the main track has an ECal PID consistent with that of a $\mu^+$ instead of a $\pi^+$.
    \item \texttt{AddTrk}: Consists of signal-like events with an additional reconstructed track contained in FGD1.
    \item \texttt{TPID}: Consists of signal-like events where the main track is not compatible with a $\pi^+$ according to TPC PID information.   
\end{itemize}

The expected and observed event rates in all samples are summarized in Table~\ref{tab:total_events}. In all samples GENIE, NEUT and data follow similar distributions both in the selected track reconstructed momentum and its angle. A deficit of data events is observed in the forward-most bin in the \texttt{EPID} sample, a feature that has been observed in the past in CC0$\pi$ studies, e.g. Refs.~\cite{MINERvA:2015ydy, MINERvA:2018hqn, T2K:2020sbd, T2K:2020jav}, and that is attributed to a mis-modeling of neutrino interactions with low squared momentum transfer, $Q^2$. As this is a known feature, we tested the robustness of the signal cross section extraction to $Q^2$ modifications and no bias was observed. Further details are later presented in  Sec.~\ref{sec:model_indep}.

\begin{table}[hbt!]
    \footnotesize
    \centering
    \caption{Total number of expected selected events in the Signal Region (SR) and in the \texttt{EPID}, \texttt{AddTrk} and \texttt{TPID} samples for NEUT and GENIE  normalized to the data POT}.
\begin{tabular}{l|rrrr}
\hline \hline
Events &  SR  & \texttt{EPID} & \texttt{AddTrk} & \texttt{TPID}  \\
\hline \hline
NEUT & 470.3  & 575.8 & 90.0 & 584.4 \\
GENIE & 492.0  & 589.3 & 95.0 & 549.6 \\
Data & 492  & 540     & 69 & 548   \\
\hline \hline
\end{tabular}
    \label{tab:total_events}
\end{table}

\section{Signal Extraction and Model Uncertainties}

The NC1$\pi^+$ cross section is measured using an unregularized binned maximum likelihood fit, the same method described in earlier T2K measurements, e.g. in Refs.~\cite{T2K:2023qjb,T2K:2023xlh}. 

\subsection{Binned Likelihood Fit}
\label{sec:bin_lik_fit}

\begin{figure}[ht!]
\centering
\includegraphics[width=0.49\textwidth]{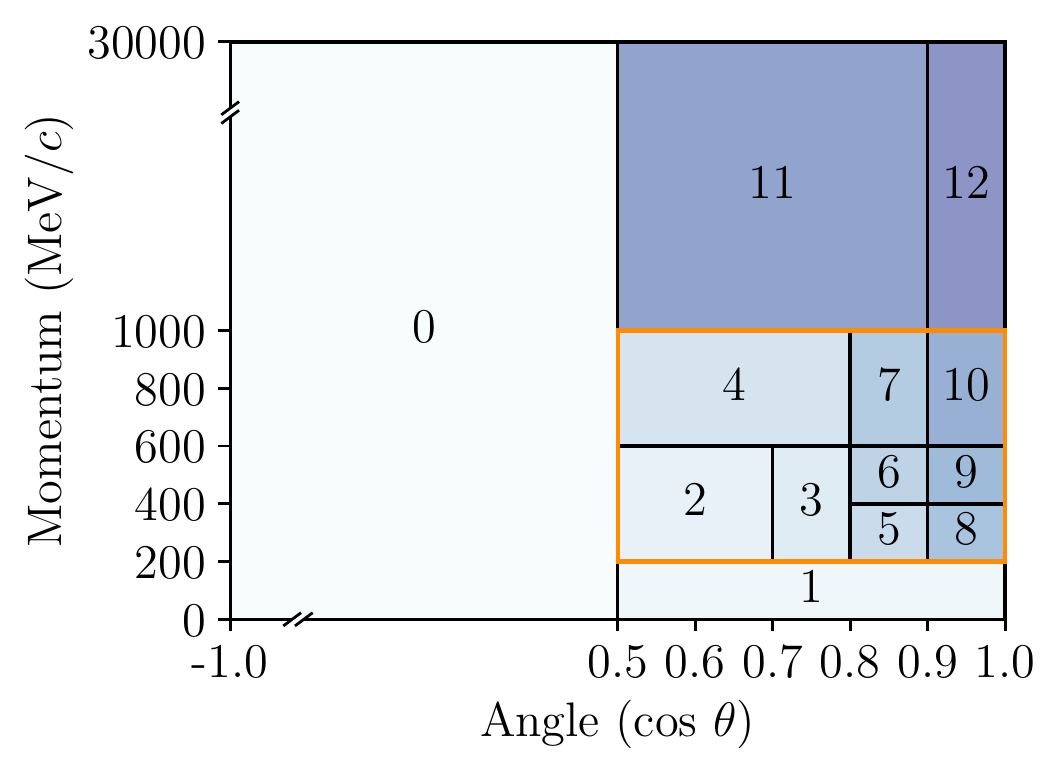}

\caption{Scheme of the true kinematic bins where the cross section is measured. Bin IDs are shown within their associated kinematic regions. The region of interest, enclosed by an orange solid line, corresponds to bins 2-10. Color shades are used to aid visualization but have no physical meaning. The figure is reproduced from Ref.~\cite{Abe:2025wxc}.}

\label{fig:2d_Bins_scheme}
\end{figure}

The signal and control samples are binned in reconstructed track momentum and angle. The expected event rate per bin is calculated by varying the nominal predictions according to model uncertainties. These variations account for uncertainties in the flux, detector and cross-section parameters. Additionally, the NC1$\pi^+$ signal cross section is adjusted using one free parameter per bin, referred to as template parameters. The binning scheme, detailed in Appendix~\ref{sec:binning}, consists of 13 (85) bins in true (reconstructed) track angle and momentum. The true kinematic bins, where the cross section is extracted, are shown in Fig.~\ref{fig:2d_Bins_scheme}. The parameters are simultaneously optimized to maximize agreement with data in both signal and background samples. This optimization provides a data-driven background constraint while unfolding detector effects and estimating the number of signal events in each bin.

The best-fit parameters and their uncertainty are found by minimizing the negative log-likelihood ratio, approximated by the chi-square:
\begin{align}
\label{eq:chi2def}
    \chi^2 \approx -2 \log \mathcal{L} = -2 \log \mathcal{L}_{\mathrm{stat}} -2 \log \mathcal{L}_{\mathrm{syst}}
\end{align}
The statistical term of the likelihood, $\mathcal{L}_{\mathrm{stat}}$, is the modified statistical log-likelihood ratio following the Barlow-Beeston method~\cite{Barlow:1993dm, Prosper:2011zz}. The systematic term of the likelihood, $\mathcal{L}_{\mathrm{syst}}$ corresponds to
\begin{align}
    \mathcal{L}_{\mathrm{syst}} = (\vec{p}-\vec{p}_{\mathrm{prior}}) V^{-1}_{\mathrm{syst}}
    (\vec{p}-\vec{p}_{\mathrm{prior}}),
\end{align}
where $\vec{p}$ represents the fit parameters under evaluation, $\vec{p}_{\mathrm{prior}}$ defines their nominal value and $V_{\mathrm{syst}}$ is the covariance matrix encoding their constraints including correlations between each other. Details on the model parameters are presented in the next section. Template parameters, used to extract the signal cross section, are unconstrained.

\subsection{Systematic Uncertainties}
\label{sec:syst_unc}
Three types of model variations are considered.

\subsubsection{Flux Model Uncertainty}
All events are grouped in bins of flavor ($\nu_e$, $\nu_\mu$, $\bar{\nu}_e$, $\bar{\nu}_\mu$) and true neutrino energy (20 bins per flavor). There is one fit parameter per flux bin, for a total of 80 flux parameters. These apply identical scaling weights to all associated events. Uncertainties across all flux bins are correlated and characterized using a flux covariance matrix produced by simultaneously varying underlying flux model parameters in the simulation according to their prior uncertainties. For the $\nu_\mu$ flavor prediction, contributing to the vast majority of events, the uncertainty is about 5\%~\cite{Vladisavljevic:2018prd} for the most important kinematic neutrino energy region, i.e.\ around the flux peak at E$_\nu\sim$~0.6 GeV.

\subsubsection{Detector Model Uncertainty}

One parameter is associated to every reconstructed kinematic bin, for a total of 85 bins summarized in Tables~\ref{tab:sig_reco_binning},~\ref{tab:sb1_reco_binning},~\ref{tab:sb2_reco_binning} and~\ref{tab:sb3_reco_binning} in Appendix~\ref{sec:binning}. Each detector parameter consists of a normalization weight that equally scales all signal and background events in that reconstructed bin. The prior error on each detector parameter is characterized by a detector covariance matrix calculated for this study using an ensemble of 500 distributions of the events selected obtained by varying simultaneously all detector effects based on their prior uncertainties (e.g. PID variables, reconstruction efficiencies, pion and proton secondary interactions, amount of pile-up, etc). Therefore, the final event rate variations reflect changes in the selection efficiency due to changes in variables used to define the selection criteria, and account for the migration of the selected track kinematics. For the bins in the signal sample the integrated rate uncertainty in the RoI is 4.5\%, dominated by the modeling of pion secondary interactions (SI) (2.0\%), the ECal PID (1.9\%), and the TPC PID (1.9\%). The overall level of uncertainty is also below 5\% for the \texttt{EPID} and \texttt{TPID} samples, with the ECal PID (TPC PID) being the dominant uncertainty for the former (latter) at the level of 2.8\% (3.0\%). For the \texttt{AddTrk} sample the integrated uncertainty is of 8.6\%, dominated by the modeling of pion SI (3.7\%) and the main track momentum resolution (3.5\%).

\subsubsection{Cross-Section Model Uncertainty}
A total of 28 cross section model parameters are considered. Unless otherwise specified their prior uncertainties and modeling match the description in Ref.~\cite{T2K:2023qjb}.

\begin{itemize}
\setlength\itemsep{-0.2em}
    \item Three parameters to vary quasi-elastic interactions with one or two nucleons ($M_{A}^{\mathrm{QE}}$, 2p2h shape, 2p2h normalization) and three parameters to account for the modeling  of resonant interactions:  $M_{A}^{\mathrm{RES}}, C^5_A$ and $I^{\mathrm{RES}}_{12}$.

    \item Five parameters to modify CC events with one (CC1$\pi$ E$_\nu<$ 2.5~GeV normalization, CC1$\pi$ E$_\nu>$ 2.5~GeV normalization) and more pions (Multi-$\pi$ normalization, Deep Inelastic Scattering (DIS) normalization, DIS shape).
    
    \item Two normalization parameters to scale neutrino neutral-current elastic (NCE) and resonant $\pi^0$ interactions with a prior uncertainty of 30\% and another to change the normalization of CC coherent pion production with a 100\% prior uncertainty.
    
    \item Three normalization parameters to scale out-of-FGD1 fiducial volume events in three independent groups split by parent type: neutrino, neutron or other; each with a prior uncertainty of 25\%.
    
    \item  Five low $Q^2$ normalization parameters, to modify CC Quasi-Elastic (CCQE) events in five $Q^2$ regions. These parameters add flexibility to the model to account for the known discrepancies observed in previous CCQE studies. These parameters and their uncertainties are analogous to those used in the most recent oscillation analysis by T2K~\cite{T2K:2023smv}.
    
    \item  Six FSI parameters related to inelastic interactions, pion absorption, production and charge exchange below and above 0.5 GeV.
    
\end{itemize}

\subsection{Cross Section Extraction and Error Propagation}
\label{sec:xsec_extraction}
The result of the fit consists of the collection of parameter values that minimize the $-2\log \mathcal{L}$ and their corresponding post-fit covariance matrix. To calculate the NC$1\pi^+$ cross section, a thousand sets of correlated parameter values are sampled from the post-fit distribution. The expected signal events in every true bin for each parameter set are used to calculate an ensemble of cross-section values by means of
\begin{align}
\label{eq:xsec_formula}
\frac{d^2\sigma_i}{dp\,d\cos\theta} = \frac{N^{\text{signal}}_i}{\varepsilon_i \Phi N^{\textup{FV}}_{\textup{nucleons}}\Delta x_i},
\end{align}
where the index \textit{i} refers to the bin number, $\sigma$ denotes the cross section, $N^{\text{signal}}$ are the number of expected signal events for the parameter set under consideration, $\varepsilon$ is the efficiency calculated from the ratio of weighted pre-selection and post-selection true signal events, $\Phi$ is the total neutrino and antineutrino flux across all flavors, adjusted by post-fit flux parameters, $\Delta x$ is the bin area, and $N^{\textup{FV}}_{\textup{nucleons}}=4.977\times 10^{29}$ is the number of nucleons in the FGD1 fiducial volume, with a corresponding uncertainty of 0.67\% included in the calculation.

\section{Measurement Validations}
\label{sec:model_indep}

\subsection{Model Variations}
\label{sec:robustness}

To validate the measurement robustness to plausible model variations, the signal cross section is extracted in controlled conditions. Each test consists of performing the measurement for a known modified event rate prediction, commonly referred to as Fake Data (FD), allowing to quantify its bias under known input conditions. The list of FD studies, labeled with unique IDs, are  presented in Table~\ref{tab:FDS_summary}. The results are as follows:
\begin{itemize}
    \item Biases of at most 5\% the uncertainty are observed when performing normalization shifts the most important backgrounds for the analyses, including variations of OOFV, CC$1\pi$, $\bar{\nu}_{\mu}$~CC, NCX$\pi^0$, CCQE, CC0$\pi$, NCE, NC-other and NC1$\pi$ with protons above 200 MeV/$c$, corresponding to IDs 0-8. Particularly relevant is the result of increasing the weight of NC1$\pi^+$ background events with protons, which supports the signal definition choice earlier described in Sec.~\ref{sec:pr_threshold}. The applied Np increase ($\times5)$ corresponds to the prediction difference between NEUT and GENIE for NC$1\pi^+$ events with protons regardless of their momentum. It is worth noting that an analogous test with an alternative signal definition including exclusively protonless NC$1\pi^+$ events yielded a bias of approximately 30\% of the measurement uncertainty, demonstrating the importance of an adequate signal definition.
    
    \item Scaling tests to check the impact of a varied normalization for the events out of the RoI ($\cos\theta_{\pi^{+}}>$0.5 and $0.2<p_{\pi^+}< 1.0$~GeV/$c$), were we do not report the measurement due to low selection purity and efficiency, translate into biases of about 10\% of the magnitude of the assessed uncertainty, and are associated to IDs 9 and 10.
    
    \item Variations of background events with low (VA$_{5\times 5}<250$) and high (VA$_{5\times 5}>250$) vertex activity, corresponding to IDs 11 and 12, yield biases of about 5\% of the magnitude of the uncertainty. The VA$_{5\times 5}>$ variable denotes the summed light yield in a 5$\times$5 volume of FGD1 bars around the reconstructed vertex, and therefore this test checks for the impact of mis-modeling in the predictions of low-momentum hadrons.
    
    \item A re-scaling of the CCQE events based on their true $Q^2$ value was done in three ways to check the impact of $Q^2$-dependent mis-modeling hinted by several experiments: 1) an overall normalization of all events with low $Q^2$, i.e.\ all those events below 0.25~GeV$^2$/c$^2$; 2) a parametric modification of those $Q^2$ events using  5 normalization parameters for $Q^2\in[0,0.25]$~GeV$^2$/c$^2$ taken in steps of $0.05$~GeV$^2$/c$^2$, and with values corresponding to the pre-fit and 3) post-fit values for those model parameters used in the oscillation analysis of T2K in Ref~\cite{T2K:2023smv}. In all cases observed biases are about 5\% of the measurement uncertainty. These tests have associated IDs 13-15.
    
    \item Positive and negative normalization shifts of the signal cross section, corresponding to IDs 16 and 17, lead to negligible biases, confirming the adequate fitter behavior to capture signal variations.

    \item A fit to the GENIE prediction scaled to the data POT,  shown in Figs.~\ref{fig:signal}~and~\ref{fig:control_regions}. GENIE event rates differ significantly with NEUT in most of its background predictions, presented in Table.~\ref{tab:events_in_signal_sample_by_topo}. An extracted cross section bias of about 15\% of the measurement uncertainty is observed. This test is the only one using statistically independent samples, therefore resulting in larger post-fit $\chi^2$. The post-fit $\chi^2$ can be sub-divided in $\chi^2_{stat}=10.2$ and $\chi^2_{syst}=1.4$. This value, as expected, is much lower than the number of degrees of freedom in the fit, since GENIE event rates are scaled to data POT but have a much smaller statistical uncertainty due to an order of magnitude larger POT in the GENIE simulation than in data.
\end{itemize}

While Table~\ref{tab:FDS_summary} summarizes the results in terms of the integrated cross-section values, we also looked for indications of bias in differential bins for each study. The observed biases in all cases are a small fraction of the measurement uncertainty supporting the flexibility of the fitted model and the adequacy of the size of the prior uncertainties.

\begin{table}[hbt!]
\small
\centering
\caption{Summary of fake data studies results. All expectations use Monte Carlo simulations normalized to the data POT. The "Ratio" column shows the ratio of the integrated cross section from each study to that obtained from an Asimov fit using nominal NEUT event rates. The post-fit $\chi^2$ corresponds to Eq.~\ref{eq:chi2def} at the best-fit-point.}
\begin{tabular}{l|llr}
\hline \hline 
ID & Description & Ratio  & Post-Fit $\chi^2$ \\
\hline \hline 
0 & OOFV$\times1.3$ & 1.00 $\pm$ 0.22 & 1.0\\
1 & CC1$\pi\times0.8$ & 1.00 $\pm$ 0.22 & 0.1\\
2 & $\bar{\nu}_\mu$~CC$\times1.2$ & 1.0 $\pm$ 0.22 & 0.9\\
3 & NCX$\pi^0\times1.2$ & 1.01 $\pm$ 0.22 & 0.1\\
4 & CCQE $\times 1.3$ & 0.99 $\pm$ 0.21 & 0.2\\
5 & CC0$\pi$ $\times 1.2$ & 1.00 $\pm$ 0.22 & 0.7\\
6 & NCE $\times 1.3$ & 1.00 $\pm$ 0.22 & 0.4\\
7 & NC-other $\times 1.3$ & 0.99 $\pm$ 0.22 & 0.0\\
8 & NC1$\pi^+$ + Np $\times 5$ & 1.01 $\pm$ 0.22 & 0.4\\
9 & $\cos\theta < 0.3 \times 0.5$ & 1.02 $\pm$ 0.22 & 1.4\\
10 & $p < 0.3$ (MeV/$c$) $\times 0.5$ & 1.02 $\pm$ 0.22 & 0.7\\
11 & BGK VA$_{5\times5}<$250 $\times 1.2$ & 1.01 $\pm$ 0.22 & 1.4\\
12 & BGK VA$_{5\times5}>$250 $\times 1.2$ & 0.99 $\pm$ 0.21 & 0.2\\
13 & Low $Q^2$ $\times 0.8$ & 0.99 $\pm$ 0.21 & 0.7\\
14 & Low $Q^2$ pre-fit OA2020~\cite{T2K:2023smv} & 0.99 $\pm$ 0.22 & 1.9\\
15 & Low $Q^2$ post-fit OA2020~\cite{T2K:2023smv} & 1.00 $\pm$ 0.22 & 0.2\\
16 & signal$\times0.8$ & 1.00 $\pm$ 0.24 & 0.0\\
17 & signal$\times1.2$ & 1.00 $\pm$ 0.20 & 0.0\\
18 & GENIE 2.8.0 & 1.03 $\pm$ 0.21 & 11.6\\
\hline \hline 
\end{tabular}
\label{tab:FDS_summary}
\end{table}

\subsection{Coverage}

\begin{figure}[ht!]
\centering
\includegraphics[width=0.49\textwidth]{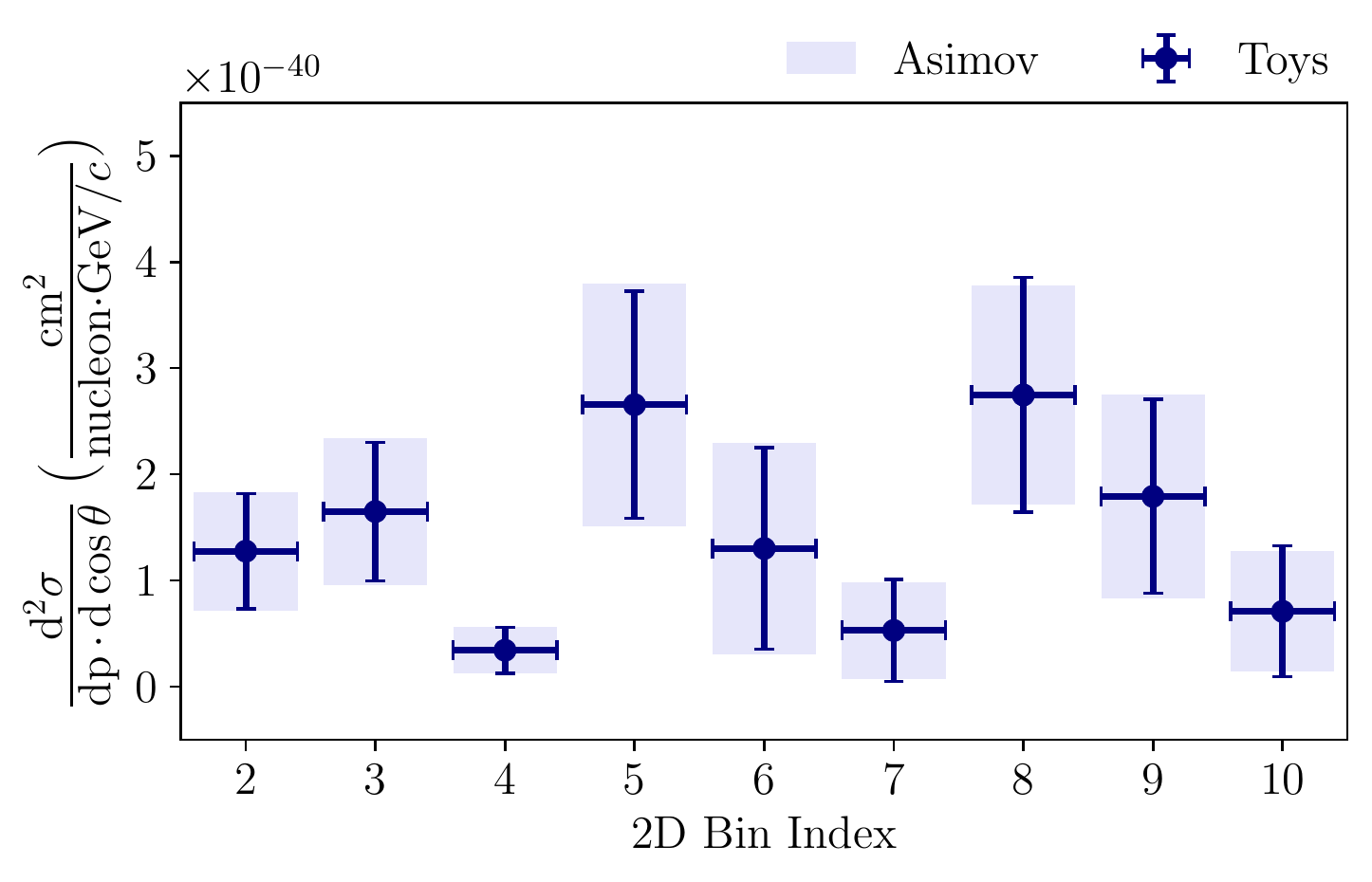}
\caption{Comparison of the differential cross section bin means and errors calculated from the post-fit Hessian matrix of a single Asimov fit to the nominal model, and the observed standard-deviation in every bin in an ensemble of 550 toys with simultaneous statistical and systematic variations according to the prior uncertainties.}
\label{fig:coverage}
\end{figure}

\begin{figure}[ht!]
\centering
\includegraphics[width=0.49\textwidth]{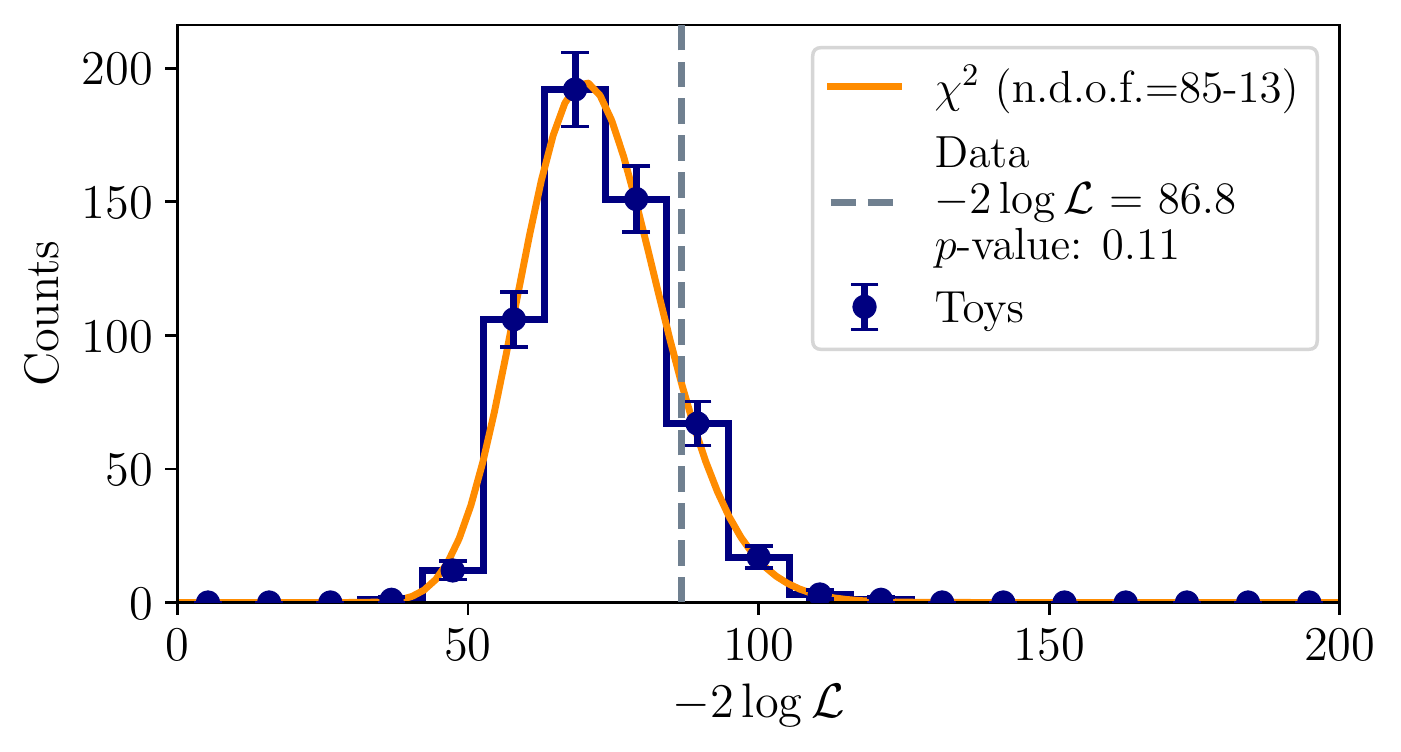}
\caption{Observed post-fit $\chi^2$ distribution in 550 toy fits compared to the theoretical $\chi^2$ expectation for the $85-13$ degrees of freedom in the fit.}
\label{fig:chi2}
\end{figure}

The coverage for the measurement was studied by performing fits to an ensemble of varied model predictions (Toys) consisting of simultaneous random statistical and systematic variations of the prediction in accordance with the pre-fit uncertainties. Statistical (systematic) variations used a Poisson (Gaussian) distribution.

In Figure~\ref{fig:coverage} the extracted double-differential cross section from a single fit to the nominal model predictions, so-called Asimov fit~\cite{Cowan:2010js}, is compared to the distribution of extracted cross sections for an ensemble of 550 toys. Good agreement is observed both for the means and the error bars for all cross section bins, supporting their Gaussian interpretation.

Figure~\ref{fig:chi2} illustrates the distribution of the post-fit $\chi^2$ for all toys, in good agreement with the theoretical expectation of a $\chi^2$ distribution with 72 degrees of freedom, corresponding to 85 bins in reconstructed kinematic space minus 13 free unconstrained parameters.

\section{Cross Section Result}
Data unbinding was performed in a two-stage process. Firstly, the pre-fit event rate in the sidebands was checked to be in qualitative agreement with data and a fit to the side-bands using nominal predictions for the signal sample was performed. No nuisance parameters (detector, flux or cross-section) were observed to be in significant tension with their nominal values. Secondly, the signal sample was also unblinded, leading to the final data fit and result that we report next.\\
The data fit has a $-2 \log \mathcal{L} = 86.8$, divided in  $-2 \log \mathcal{L}_{stat} = 75.9$ and $-2 \log \mathcal{L}_{syst} = 10.9$. The associated p-value is 0.11, supporting a statistically reasonable goodness-of-fit. The largest contribution to $-2 \log \mathcal{L}$ is the statistical term, which is in a similar level of agreement in the signal sample and the sidebands. In particular, the statistical $\chi^2/\text{nbins}$ in the signal, \texttt{EPID}, \texttt{AddTrk} and \texttt{TPID} samples corresponds to: $23.2/33$, $25.9/23$, $3.8/6$, and $22.9/23$. \\

\begin{figure}[ht!]
\centering
\includegraphics[width=0.49\textwidth]{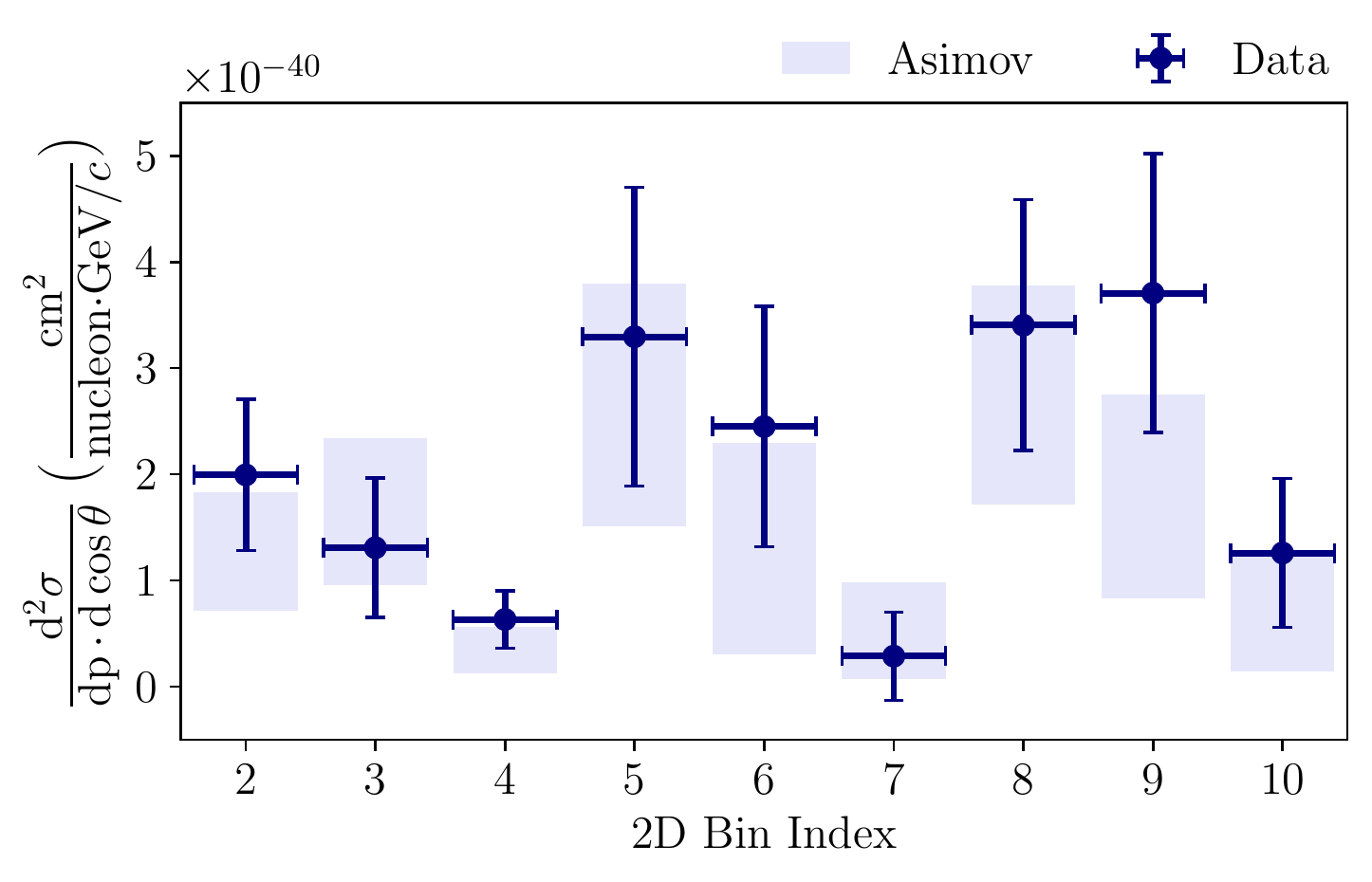}

\caption{Differential cross section from NEUT~v5.4.0 compared to the measured differential data result.}
\label{fig:data_diff}
\end{figure}

\begin{figure}[ht!]
\centering
\includegraphics[width=0.49\textwidth]{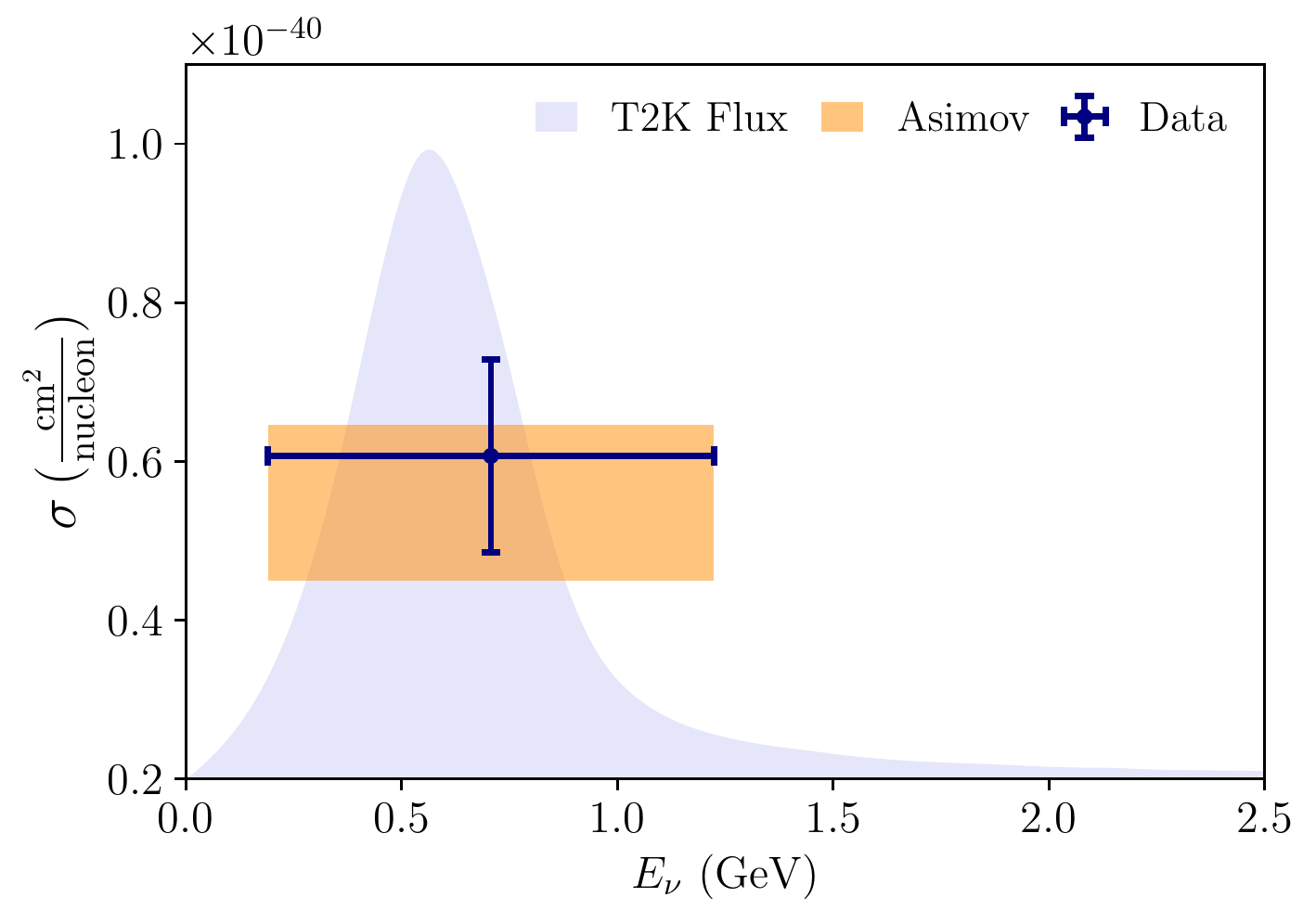}

\caption{Integrated cross section from NEUT~v5.4.0 compared to the measured integrated data result. The x-axis error bar covers one standard deviation of the T2K neutrino flux around its mean energy of 0.70 GeV.}
\label{fig:data_integ}
\end{figure}

The measurement of the NC1$\pi^+$ double-differential cross section without measured protons above 200 MeV/$c$ is presented in Fig.~\ref{fig:data_diff}. Bins 5, 6, 8 and 9, covering the kinematic region with highest purity and signal abundance all prefer a higher signal cross section than in NEUT.

The flux-averaged integrated cross section for the signal, presented in Fig.~\ref{fig:data_integ}, in the RoI is:
\begin{align*}
    \sigma = (6.07 \pm 1.22 ) \times 10^{-41} \,\, \text{cm}^2/\text{nucleon}\,.
\end{align*}
The result translates into a central value for the cross-section 35\% larger than the one expected by NEUT, corresponding to a 1.3$~\sigma$ pull.

Additional robustness tests on the final result are provided in Appendix~\ref{sec:additional_tests}.

\section{Conclusions}
T2K has identified the largest sample of data associated to NC1$\pi^+$ interactions, and used it to perform the first differential cross-section measurement of this process. The unique design of T2K's near detector ND280, utilizing a 0.2~T magnetic field and diverse sub-detector technologies, has proved essential in realizing this study. ND280's magnet and the TPCs were crucial to separate $\mu^-$ from $\pi^+$ and to precisely measure the track kinematics. FGD1 distinguished forward- from backward-going single $\pi^+$-like tracks, while the ECal separated $\mu^+$ from $\pi^+$. Thanks to these combined performances and the large amount of data collected by ND280, the signal selection criteria presented in this article were capable of identifying over two hundred expected signal events. In the RoI of the signal sample, an efficiency of about 30\% and a purity of 50\% was achieved. 

The dominant background in this analysis, $\bar{\nu}_\mu$~CC interactions, stems from the $\bar{\nu}$ contamination when running in $\nu$-beam mode. Therefore, while a similar measurement of the NC1$\pi^-$ cross section could be attempted with ND280 in the future using $\bar{\nu}$-beam mode data, the relatively higher contamination of $\nu$ would result, in principle, in substantially lower signal purity for the selected signal sample. Nevertheless, the success in isolating a high-purity sample of wrong-sign $\bar{\nu}_\mu$~CC events presented in this article, motivates further exploration on the ability of T2K to measure NC1$\pi^-$ interactions.

The signal cross section was extracted using an unregularized binned likelihood fit following the same method utilized in other recent T2K cross section measurements. Validations specific to this analysis have been performed using sets of test fake data with controlled variations to quantify the measurement bias. In all cases a small or negligible bias was observed supporting the robustness of the cross section extraction method. Additional studies support the Gaussian interpretation of the measurement uncertainties.

The~extracted double-differential cross section shows a weak preference for larger values than NEUT, particularly in regions with a very forward-going $\pi^+$. The flux-averaged integrated cross section is 35\% larger than NEUT, corresponding to a 1.3~$\sigma$ pull.

The measurement results have an associated data release that can be found in Ref.~\cite{T2K_release}.

\color{black}

\section*{Acknowledgments}
The T2K collaboration would like to thank the J-PARC staff for superb accelerator performance. We thank the CERN NA61/SHINE Collaboration for providing valuable particle production data. We acknowledge the support of MEXT,   JSPS KAKENHI (JP16H06288, JP18K03682, JP18H03701, JP18H05537, JP19J01119, JP19J22440, JP19J22258, JP20H00162, JP20H00149, JP20J20304, JP24K17065) and bilateral programs (JPJSBP120204806, JPJSBP120209601),  Japan; NSERC, the NRC, and CFI, Canada; the CEA and CNRS/IN2P3, France; the Deutsche Forschungsgemeinschaft (DFG, German Research Foundation) 397763730, 517206441, Germany; the NKFIH (NKFIH 137812 and TKP2021-NKTA-64), Hungary; the INFN, Italy; the Ministry of Science and Higher Education (2023/WK/04) and the National Science Centre (UMO-2018/30/E/ST2/00441, UMO-2022/46/E/ST2/00336 and UMO-2021/43/D/ST2/01504), Poland;  the RSF (RSF 24-12-00271) and the Ministry of Science and Higher Education, Russia; MICINN  (PID2022-136297NB-I00 /AEI/10.13039/501100011033/ FEDER, UE, PID2021-124050NB-C31, PID2020-114687GB-I00,  PID2019-104676GB-C33), Government of Andalucia (FQM160, SOMM17/6105/UGR) and the University of Tokyo ICRR's Inter-University Research Program FY2023 Ref. J1, and ERDF and European Union NextGenerationEU funds (PRTR-C17.I1) and CERCA program, and University of Seville grant Ref. VIIPPIT-2023-V.4, and Secretariat for Universities and Research of the Ministry of Business and Knowledge of the Government of Catalonia and the European Social Fund (2022FI\_B 00336), Spain; the SNSF and SERI (200021\_185012, 200020\_188533, 20FL21\_186178I), Switzerland; the STFC and UKRI, UK; the DOE, USA; and NAFOSTED (103.99-2023.144,IZVSZ2.203433), Vietnam. We also thank CERN for the UA1/NOMAD magnet, DESY for the HERA-B magnet mover system, the BC DRI Group, Prairie DRI Group, ACENET, SciNet, and CalculQuebec consortia in the Digital Research Alliance of Canada, and GridPP in the United Kingdom, and the CNRS/IN2P3 Computing Center in France and NERSC (HEP-ERCAP0028625). In addition, the participation of individual researchers and institutions has been further supported by funds from the ERC (FP7), “la Caixa” Foundation  (ID 100010434, fellowship code LCF/BQ/IN17/11620050), the European Union’s Horizon 2020 Research and Innovation Programme under the Marie Sklodowska-Curie grant agreement numbers 713673 and 754496, and H2020 grant numbers  RISE-GA822070-JENNIFER2 2020 and RISE-GA872549-SK2HK; the JSPS, Japan; the Royal Society, UK; French ANR grant number ANR-19-CE31-0001 and ANR-21-CE31-0008; and  Sorbonne Université Emergences programmes; the SNF Eccellenza grant number PCEFP2\_203261;  the VAST-JSPS (No. QTJP01.02/20-22);  and the DOE Early Career programme, USA. For the purposes of open access, the authors have applied a Creative Commons Attribution license to any Author Accepted Manuscript version arising.

\bibliographystyle{apsrev4-1}
\bibliography{biblio}

\clearpage

\appendix

\section{Additional Robustness Tests}
\label{sec:additional_tests}

We show in Fig.~\ref{fig:alternative_xsec_model} the observed variations in the final cross-section results when using alternative cross-section model parameters during the unfolding. The alternative tests are as follows:
\begin{enumerate}\setlength\itemsep{-0.em}
    \item \textbf{Disabled:}  We switch off completely the cross-section model parameters. Namely, we only use detector and flux parameters during unfolding.
    \item \textbf{Norm-Only:}  All shape parameters ($M_{A}^{\mathrm{QE}}$, 2p2h shape, $M_{A}^{\mathrm{RES}}, C^5_A$ and $I^{\mathrm{RES}}_{12}$, and FSI parameters) are switched off.
    \item \textbf{Low-Q$^2$:}  All the low-$Q^2$ normalization parameters are switched off.
    \item \textbf{No OOFV:}  All the OOFV normalization parameters are switched off.
\end{enumerate}
Even in the most extreme case, where the cross-section model is fully disabled, the alternative result is well within one sigma of the reported cross-section result for each differential bin. In all cases the pull between the NEUT 5.4.0 prediction and the final result is much larger than the observed variations for the alternative cross-section models, supporting the robustness of the result. Furthermore, fits where certain sub-sets of cross-section models are disabled (Norm-Only, Low-$Q^2$, No OOFV), which directly affect the most abundant categories of selected events in the signal sample and the sidebands, lead to minimal changes in the reported result. 
\begin{figure}[ht!]
\centering
\includegraphics[width=0.49\textwidth]{Figures/xsec_model_additional_validations.pdf}
\caption{Differentiable cross section for NEUT 5.4.0 and the data result, as well as the data results for alternative cross-section model parametrization choices.}
\label{fig:alternative_xsec_model}
\end{figure}
\vfill

\section{Binning}
\label{sec:binning}
\vspace{-0.7cm}
\renewcommand{\arraystretch}{0.85}
\begin{table}[H]
    \footnotesize
    \centering
    \caption{Binning in true kinematic space. Bins 0, 1, 11 and 12 cover the kinematics space out of the region of interest.}
\begin{tabular}{r|r|r|r|r}
\toprule \toprule
Index & min $\cos{\theta}$ & max $\cos{\theta}$ & min p (MeV/$c$) & max p (MeV/$c$) \\
\hline \hline
0 & -1 & 0.5 & 0 & 30000	   \\
1 & 0.5 & 1 & 0 & 200	       \\
2 & 0.5 & 0.7 & 200 & 600	   \\
3 & 0.7 & 0.8 & 200 & 600	   \\
4 & 0.5 & 0.8 & 600 & 1000	   \\
5 & 0.8 & 0.9 & 200 & 400	   \\
6 & 0.8 & 0.9 & 400 & 600	   \\
7 & 0.8 & 0.9 & 600 & 1000	   \\
8 & 0.9 & 1.0 & 200 & 400	   \\
9 & 0.9 & 1.0 & 400 & 600	   \\
10 & 0.9 & 1.0 & 600 & 1000	   \\
11 & 0.5 & 0.9 & 1000 & 30000  \\
12 & 0.9 & 1.0 & 1000 & 30000  \\
\hline \hline
\end{tabular}
    \label{tab:true_binning}
\end{table}
\vspace{-10pt}

\begin{table}[H]
    \footnotesize
    \centering
    \caption{Binning in reconstructed kinematic space for the signal sample.}
\begin{tabular}{r|r|r|r|r}
\hline \hline
Index & min $\cos{\theta}$ & max $\cos{\theta}$ & min p (MeV/$c$) & max p (MeV/$c$) \\
\hline \hline
0 & -1.0 & 0.5 & 0 & 30000	   \\
1 & 0.5 & 0.8 & 0 & 200	       \\
2 & 0.8 & 1.0 & 0 & 200	       \\
3 & 0.5 & 0.6 & 200 & 600	   \\
4 & 0.6 & 0.7 & 200 & 350	   \\
5 & 0.6 & 0.7 & 350 & 600	   \\
6 & 0.5 & 0.8 & 600 & 1000	   \\
7 & 0.5 & 0.8 & 1000 & 30000   \\
8 & 0.7 & 0.8 & 200 & 300	   \\
9 & 0.7 & 0.8 & 300 & 400	   \\
10 & 0.7 & 0.8 & 400 & 600	   \\
11 & 0.8 & 0.9 & 200 & 300	   \\
12 & 0.8 & 0.9 & 300 & 400	   \\
13 & 0.8 & 0.9 & 400 & 500	   \\
14 & 0.8 & 0.9 & 500 & 600	   \\
15 & 0.8 & 0.9 & 600 & 800	   \\
16 & 0.8 & 0.9 & 800 & 1000	   \\
17 & 0.8 & 0.9 & 1000 & 1500   \\
18 & 0.8 & 0.9 & 1500 & 30000  \\
19 & 0.9 & 1 & 200 & 300	   \\
20 & 0.9 & 1 & 300 & 400	   \\
21 & 0.9 & 1 & 400 & 500	   \\
22 & 0.9 & 1 & 500 & 600	   \\
23 & 0.9 & 1 & 600 & 700	   \\
24 & 0.9 & 1 & 700 & 850	   \\
25 & 0.9 & 1 & 850 & 1000	   \\
26 & 0.9 & 1 & 1000 & 1200	   \\
27 & 0.9 & 1 & 1200 & 1500	   \\
28 & 0.9 & 1 & 1500 & 1700	   \\
29 & 0.9 & 1 & 1700 & 2000	   \\
30 & 0.9 & 1 & 2000 & 2400	   \\
31 & 0.9 & 1 & 2400 & 3200	   \\
32 & 0.9 & 1 & 3200 & 30000	   \\
\hline \hline
\end{tabular}
    \label{tab:sig_reco_binning}
\end{table}

\vfill
\clearpage

\begin{table}[H]
    \footnotesize
    \centering
    \caption{Binning in reconstructed kinematic space for the \texttt{EPID} sample.}
\begin{tabular}{r|r|r|r|r}
\toprule \toprule
Index & min $\cos{\theta}$ & max $\cos{\theta}$ & min p (MeV/$c$) & max p (MeV/$c$) \\
\hline \hline
33 & -1.0 & 0.5 & 0 & 30000	   \\
34 & 0.5 & 1.0 & 0 & 300	   \\
35 & 0.5 & 0.7 & 300 & 600	   \\
36 & 0.5 & 0.7 & 600 & 30000	   \\
37 & 0.7 & 0.8 & 300 & 600	   \\
38 & 0.7 & 0.8 & 600 & 800	   \\
39 & 0.7 & 0.8 & 800 & 30000	   \\
40 & 0.8 & 0.9 & 300 & 500	   \\
41 & 0.8 & 0.9 & 500 & 600	   \\
42 & 0.8 & 0.9 & 600 & 800	   \\
43 & 0.8 & 0.9 & 800 & 1000	   \\
44 & 0.8 & 0.9 & 1000 & 1300	   \\
45 & 0.8 & 0.9 & 1300 & 30000	   \\
46 & 0.9 & 1 & 300 & 500	   \\
47 & 0.9 & 1 & 500 & 600	   \\
48 & 0.9 & 1 & 600 & 700	   \\
49 & 0.9 & 1 & 700 & 800	   \\
50 & 0.9 & 1 & 800 & 1100	   \\
51 & 0.9 & 1 & 1100 & 1500	   \\
52 & 0.9 & 1 & 1500 & 2000	   \\
53 & 0.9 & 1 & 2000 & 2500	   \\
54 & 0.9 & 1 & 2500 & 3500	   \\
55 & 0.9 & 1 & 3500 & 30000	   \\
\hline \hline
\end{tabular}
    \label{tab:sb1_reco_binning}
\end{table}

\begin{table}[H]
    \footnotesize
    \centering
\caption{Binning in reconstructed kinematic space for the \texttt{AddTrk} sample.}
\begin{tabular}{r|r|r|r|r}
\hline \hline
Index & min $\cos{\theta}$ & max $\cos{\theta}$ & min p (MeV/$c$) & max p (MeV/$c$) \\
\hline \hline
56 & -1.0 & 0.6 & 200 & 30000	   \\
57 & 0.6 & 0.8 & 200 & 30000	   \\
58 & -1 & 1.0 & 0 & 200	           \\
59 & 0.8 & 1.0 & 200 & 600	       \\
60 & 0.8 & 1.0 & 600 & 1500	       \\
61 & 0.8 & 1.0 & 1500 & 30000	   \\
\hline \hline
\end{tabular}
    \label{tab:sb2_reco_binning}
\end{table}

\begin{table}[H]
    \footnotesize
    \centering
\caption{Binning in reconstructed kinematic space for the \texttt{TPID} sample.}
\begin{tabular}{r|r|r|r|r}
\hline \hline
Index & min $\cos{\theta}$ & max $\cos{\theta}$ & min p (MeV/$c$) & max p (MeV/$c$) \\
\hline \hline
62 & -1.0 & 0.6 & 0 & 30000	   \\
63 & 0.6 & 1 & 0 & 400	       \\
64 & 0.6 & 0.7 & 400 & 800	   \\
65 & 0.6 & 0.7 & 800 & 30000   \\
66 & 0.7 & 0.8 & 400 & 600	   \\
67 & 0.7 & 0.8 & 600 & 800	   \\
68 & 0.7 & 0.8 & 800 & 1300	   \\
69 & 0.8 & 0.9 & 400 & 500	   \\
70 & 0.8 & 0.9 & 500 & 600	   \\
71 & 0.8 & 0.9 & 600 & 700	   \\
72 & 0.8 & 0.9 & 700 & 800	   \\
73 & 0.8 & 0.9 & 800 & 1000	   \\
74 & 0.8 & 0.9 & 1000 & 1300   \\
75 & 0.7 & 0.9 & 1300 & 30000  \\
76 & 0.9 & 1 & 400 & 500	   \\
77 & 0.9 & 1 & 500 & 600	   \\
78 & 0.9 & 1 & 600 & 700	   \\
79 & 0.9 & 1 & 700 & 800	   \\
80 & 0.9 & 1 & 800 & 900	   \\
81 & 0.9 & 1 & 900 & 1000	   \\
82 & 0.9 & 1 & 1000 & 1200	   \\
83 & 0.9 & 1 & 1200 & 1500	   \\
84 & 0.9 & 1 & 1500 & 30000	   \\
\hline \hline
\end{tabular}
    \label{tab:sb3_reco_binning}
\end{table}

\end{document}